\documentclass[10pt,conference]{IEEEtran}
\usepackage{cite}
\usepackage{amsmath,amssymb,amsfonts}

\usepackage{algpseudocode} 
\usepackage{amsmath}
\usepackage{makecell}

\usepackage{tikz}

\usepackage{threeparttable}
\usepackage{pifont}

\usepackage{graphicx}
\usepackage{textcomp}
\usepackage{xcolor}
\usepackage{color}
\usepackage{booktabs}
\usepackage[hyphens]{url}
\usepackage{multirow} 
\usepackage{tabulary}
\usepackage{algorithm}
\usepackage{xspace}
\usepackage{hyperref}

\def\BibTeX{{\rm B\kern-.05em{\sc i\kern-.025em b}\kern-.08em
    T\kern-.1667em\lower.7ex\hbox{E}\kern-.125emX}}

\title{TENET: An Efficient Sparsity-Aware LUT-Centric Architecture for Ternary LLM Inference On Edge}
\author{
    \IEEEauthorblockN{
        Zhirui Huang\IEEEauthorrefmark{1}\IEEEauthorrefmark{2},
        Rui Ma\IEEEauthorrefmark{1},
        Shijie Cao\IEEEauthorrefmark{1},
        Ran Shu\IEEEauthorrefmark{1},
        Ian Wang\IEEEauthorrefmark{4},\\
        Ting Cao\IEEEauthorrefmark{3},
        Chixiao Chen\IEEEauthorrefmark{2},
        Yongqiang Xiong\IEEEauthorrefmark{1}
    }
    \IEEEauthorrefmark{1}\textit{Microsoft Research Asia}
    \IEEEauthorrefmark{2}\textit{Fudan University}
    \IEEEauthorrefmark{3}\textit{Tsinghua University}
    \IEEEauthorrefmark{4}\textit{Microsoft}\\
    \IEEEauthorrefmark{1}\{v-zhirhuang, mrui\}@microsoft.com
}

\begin{document}
\maketitle
\thispagestyle{plain}
\pagestyle{plain}

\newcommand{\blackding}[1]{\ding{\numexpr181+#1\relax}}

\newcommand{\nickname}{TENET\xspace}
\newcommand{\RM}[1]{[\textcolor{orange}{RM: #1}]}
\newcommand{\ZH}[1]{[\textcolor{green}{ZH: #1}]}
\newcommand{\RS}[1]{[\textcolor{cyan}{RS: #1}]}


\begin{abstract}

    Ternary quantization has emerged as a powerful technique for reducing both computational and memory footprint of large language models (LLM), enabling efficient real-time inference deployment without significantly compromising model accuracy. 
    Conventional LLM inference platforms (e.g GPUs) cannot capitalize on its benefits, as they (i) lack native support for ternary arithmetic and memory specialization and (ii) remain severely under-utilized in low-batch, real-time scenarios.
    In this work, we propose {\em \nickname}, a sparse-aware LUT-centric architecture that co-optimizes algorithm, compute, and memory for ternary LLM inference.
    To maximize the efficiency of Ternary Linear layer, \nickname introduces a Sparse Ternary LUT (STL) core that optimizes ternary mixed-precision GEMM using a symmetric precompute lookup table. It also features Dynamic Activation N:M Sparsity to exploit the sparsity within the activation of each token. 
    Additionally, we propose a LUT-based 64B:80B ternary weight decompression module to fully exploit the memory efficiency of ternary values.
    At the system level, we design a heterogeneous TENET accelerator with full programmability that integrates STL cores with high-precision cores. An associated Linear-Projection-aware Sparse Attention dataflow is introduced to optimize memory access and hardware utilization.
    We implement \nickname accelerator prototype on both FPGA and ASIC platforms. 
    Experiments across various model sizes and workloads demonstrate that \nickname-FPGA and \nickname-ASIC improve energy efficiency by 4.3$\times$ and 21.1$\times$, respectively, compared to the A100 GPU. Furthermore, \nickname-ASIC achieves a 2.7$\times$ average speedup compared to the A100 GPU in end-to-end inference latency.

\end{abstract}

\section{Introduction}

Since the astonishing success of ChatGPT-3.5 which catalyzed the AI boom in 2023, large language models (LLMs) have demonstrated outstanding performance on tasks in multiple domains \cite{black2024pi_0, qu2024llms, zhouAutoVLAVisionLanguageActionModel2025, kimSurveyIntegrationLargeRobot2024, abdinPhi3TechnicalReport2024, llmdriver}.
Beyond data centers, LLMs are now increasingly being deployed on edge devices (e.g., smart phones, AR glasses, IoT devices, robots, autonomous vehicles, etc.) to enable secure, real-time and offline ability of inference.
However, edge devices have stringent resource and power constraints. Therefore, it is crucial to develop technologies that provide both high model inference quality and speed under those restrictions.
Quantization is a key technique to improve LLM efficiency by lowering the bit-width of the model parameters. 
In essence, all quantization methods involve a trade-off, where some degree of model accuracy is sacrificed in exchange for gains in performance and energy efficiency—essentially trading the quality of intelligence for greater output capacity.


\begin{figure}
    \centering
    \includegraphics[width=\linewidth]{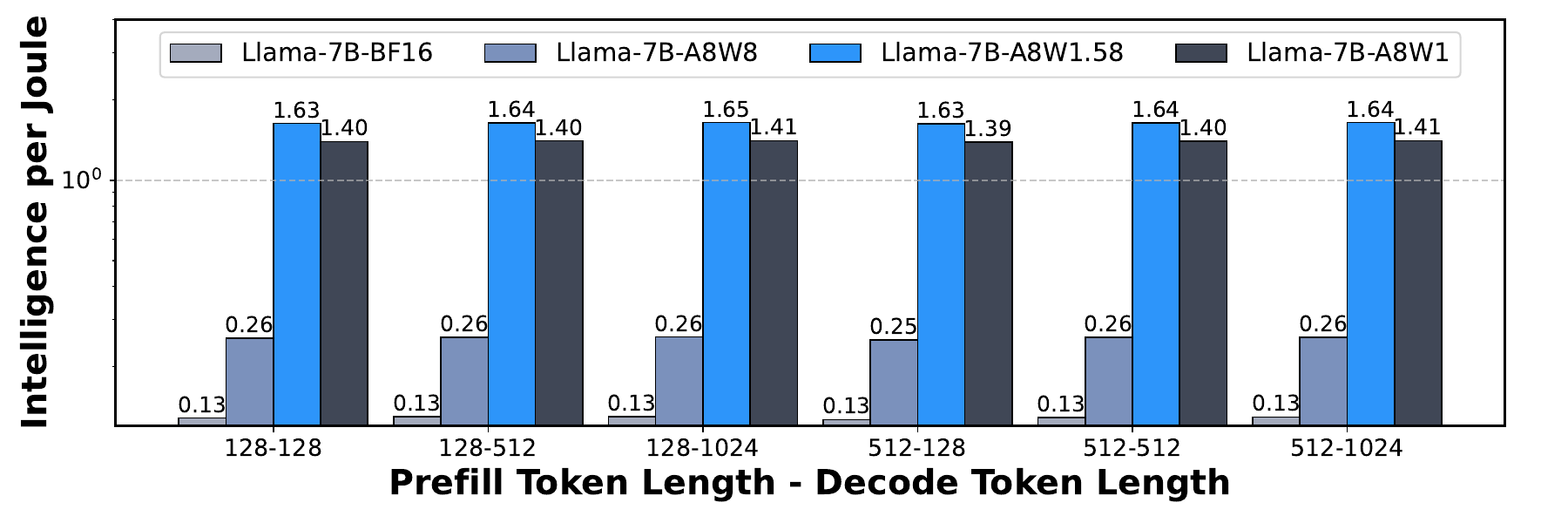}
    \vspace{-20pt}
    \caption{Compare the maximum IPJ of the Llama-7B model with different prefill-decode length under various quantization precisions, assuming weights are only loaded once at prefilling stage and once every token at decode stage}
    
    \label{fig:ipj}
\end{figure}

Therefore, we introduce \textbf{Intelligence Per Joule (IPJ)} metric, to assess a model's overall ability to generate intelligence under a fixed power budget and consistent workload condition, which is defined as:
\[IPJ=\frac{\#token}{perplexity \cdot J}=\frac{\#token/s}{perplexity \cdot W}\]
Specifically, \(1/perplexity\) quantifies the quality of the generated content\cite{TorchevalmetricsPerplexityTorchEvalMain} and \(\#tokens/J\) represents the amount of content generated within the power budget. Since \(1/perplexity\) mathematically represents the average likelihood of generating a correct token, IPJ can also be interpreted as the expected number of correct tokens generated per Joule of energy. We estimate the theoretical maximum IPJ of the LLama-7B model with various quantization precisions, assuming HBM2 as off-chip memory and 7nm logic process\cite{tpuv4i}. As shown in Figure \ref{fig:ipj}, the \textbf{ternary} quantized model achieves the best IPJ. It achieves extreme efficiency by expressing each model weight with \{-1, 0, 1\}. Meanwhile, through techniques such as Quantization-Aware Training (QAT)\cite{liuLLMQATDataFreeQuantization2023,chenEfficientQATEfficientQuantizationAware2025}, the ternary model is capable of achieving competitive model accuracy compared to full-precision models\cite{bitnet_v2,wang2023bitnet}.
Recent studies on various models (e.g., Bert\cite{ternarybert}, Llama3\cite{bitnet_v2}, Deepseek-R1\cite{deepseek-1_58}, Mamba\cite{yu-etal-2025-slender}, etc.) also demonstrate the successful adoption of ternary quantization in practice.


\begin{figure}
    \centering
    \includegraphics[width=\linewidth]{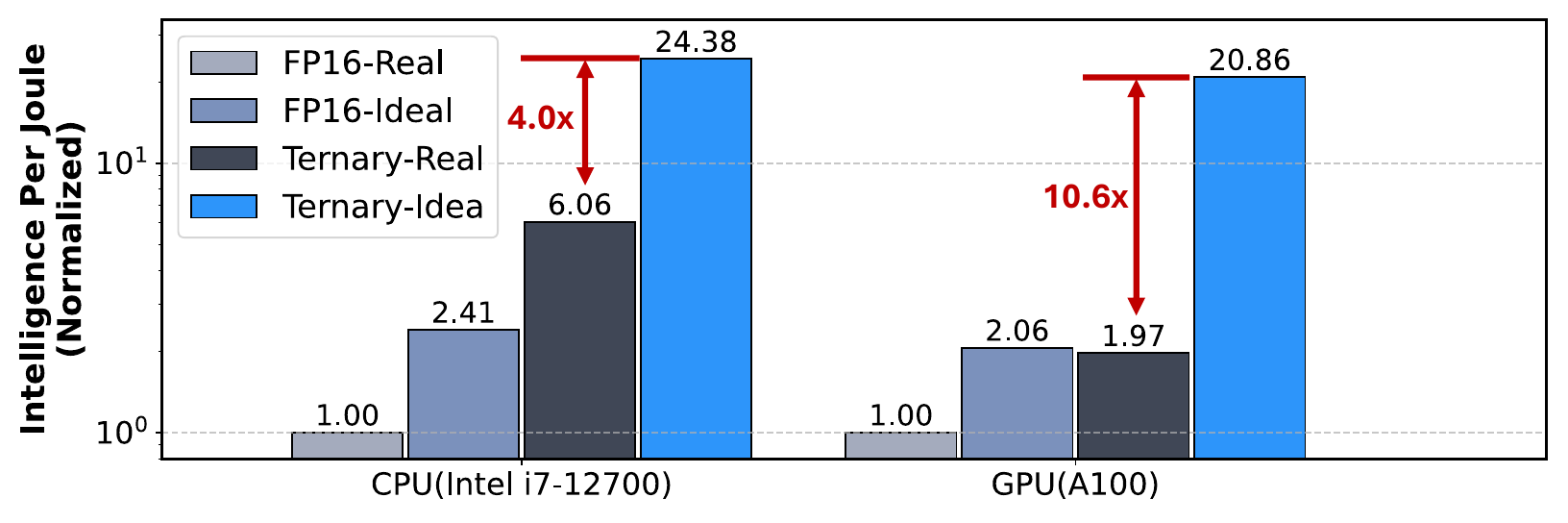}
    \vspace{-20pt}
    \caption{Achieved IPJ of running one-batch 512-512 inference of Bitnet-3B model with different weight precision on Nvidia A100-40G GPU and Intel i7-12700 CPU, compared to ideal IPJ}
    \label{fig:motivation}
    \vspace{-10pt}
\end{figure}



%



However, the efficiency of real-time inference of ternary LLMs on existing hardware is far from optimal.
We evaluated optimized implementations of one-batch inference for the Bitnet-3B \cite{wang2023bitnet} model on modern CPUs and GPUs, and compared their IPJ with a theoretical maximum that assumes linear scaling of compute throughput for ternary operations and 100\% utilization of compute and memory resources.
As shown in Figure \ref{fig:motivation}, there is a 2$\times$ gap between real and ideal IPJ of the FP16 model, indicating the difficulty to achieve high utilization on existing hardware in the one-batch scenario.
For the ternary model, although it achieves higher IPJ than the FP16 model, it is still far from ideal.
The IPJ gap relative to the ideal is 4.0$\times$ on CPU and 10.6$\times$ on GPU.
The huge gap arises because modern commodity hardware fails to fully harness the efficiency of ternary operations. From a computational perspective, they fail to exploit the inherent computation efficiency of ternary values, as ternary weights are de-quantized to align with high-precision activation\cite{flightllm,wang2024ladder}. 
From a memory perspective, the existing hardware, due to the memory alignment requirement\cite{wang2025bitnet}, often treats the ternary format as INT2, despite their theoretical minimum representation requiring only 1.58 bits.
This underscores the need for efficient acceleration of ternary LLMs on edge devices.

Additionally, as illustrated in Figure \ref{fig:bg-attention}, existing ternary LLMs primarily focus on compressing the linear weights, while leaving the attention layer still computed in high precision. 
However, as the token length grows, the unoptimized attention mechanism becomes the dominant contributor to both computation and memory access during inference, where memory efficiency is critical for energy efficiency.

To achieve efficient real-time inference on edge by fully leveraging the performance and energy efficiency advantages of Ternary LLMs, 
we propose an LUT-centric algorithm-hardware co-design, named \textbf{\nickname} by fully leveraging the intra- and inter- token sparsity and the symmetry characteristics of the ternary format.
It features a two-fold design to address the challenges through computation, memory, and architecture co-optimizations. 

\textbf{Ternary Linear Computation and Memory Optimization Design:} We introduce \blackding{1} Sparse Ternary LUT (STL) Core which exploits the symmetry of the ternary format with LUT-based computing logic.
Meanwhile, we introduce \blackding{2} Dynamic Activation N:M Sparsity to the STL Core, which leverages {\em intra-token} sparsity to reduce complexity arising from large hidden dimensions. 
For static model weights, we design a hardware-accelerated \blackding{3} LUT-based 64B$:$80B Ternary Weight Decompression (TWD) module that employs a compact packing format to represent each ternary weight with 1.6 bits, which alleviates the memory-bound issue during the decoding stage.

\textbf{Ternary LLM Inference System-level Design:} Based on STL Core, we propose the \blackding{4} TENET heterogeneous architecture, featuring a Linear-Projection-aware Sparse Attention dataflow that exploits {\em inter-token} sparsity in the attention mechanism to enable data reuse and computation fusion between QKV linear projections and attention, thereby eliminating frequent memory accesses from the attention mechanism during prefilling stage. Additionally, we develop an instruction set based on an off-the-shelf DNN accelerator\cite{brainwave} to ensure full programmability. To balance model accuracy and hardware efficiency, we propose a Design Space Exploration framework to maximize IPJ.

Algorithm experiments show that the optimized Sparse BitNet achieves comparable accuracy compared to the vanilla full precision Llama LLM with the same model size. 
We implemented \nickname accelerator on both FPGA and ASIC platforms. 
Experiments across different model sizes and workloads demonstrate that \nickname-ASIC achieves an average speedup of 27.9$\times$ over Intel 12700 CPU running SoTA bitnet.cpp framework, and 2.7$\times$ over A100 GPU with optimized kernels.
TENET-FPGA and TENET-ASIC improve energy efficiency by 4.3$\times$ and 21.1$\times$ compared to the A100 GPU, respectively. 

\section{Background and Motivation}


\subsection{Real time LLM Inference on Edge}
Deploying LLM inference on edge devices enables a new generation of intelligent, privacy-preserving applications that operate locally without relying on cloud connectivity.
For example, LLMs are deployed on personal\cite{mental-llm} or wearable \cite{odsearch} devices to monitor and analyze health conditions, without sending sensitive data to a cloud server.
Autonomous driving cars leverage on-device LLMs to understand the circumstance to help with navigation and decision-making \cite{drivevlm,llmdriver}. Low latency and connectivity independence are crucial to ensure safety.
Companion robots powered by LLM on-device \cite{3dllm} can understand and interact with the 3D world in real time.
These applications often feature low batch (even one batch) and low latency as they directly interact with one or very few users. They also demand high area and energy efficiency, as edge devices are often area/power constrained. Due to the low efficiency of GPUs and the limited performance of CPUs in low-batch LLM inference, extensive efforts from both academia \cite{flightllm,tellme,edgellm} and industry \cite{kinara-ara-2,coral,brainwave} have been made to build specialized hardware for real-time LLM inference on edge devices.

\begin{figure}[!t]
    \centering
    \includegraphics[width=8.8cm]{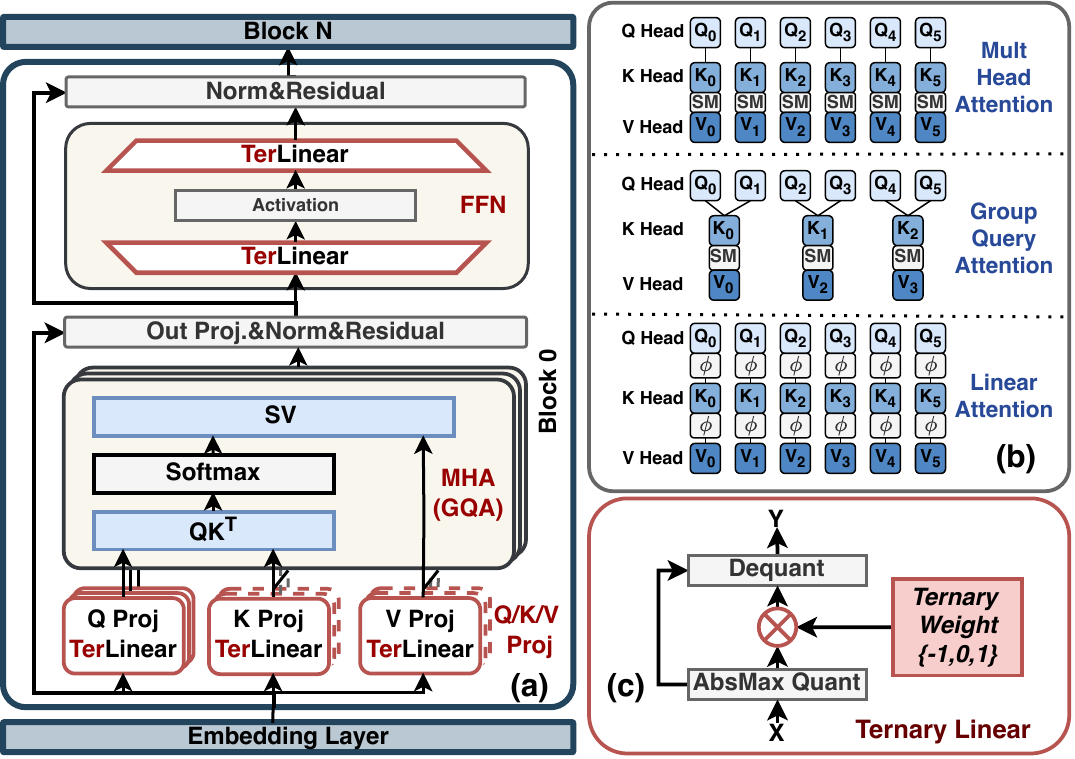}
    \caption{(a) Conventional Ternary LLM Architecture(e.g. BitNet). Ternary Linear module are colored red and high precision GEMM are blue (b) Attention mechanism variants (c) The computation flow of Ternary Linear }
    \label{fig:bg-attention}
    \vspace{-10pt}
\end{figure}

\subsection{LLM and Ternary Quantization}
\label{sec:bg_llm}
Figure \ref{fig:bg-attention}(a) shows the general architecture of a Transformer Block, which is the basic building block of modern LLMs. It consists of three layers. The \(TL\)-token input sequence, represented as embedding matrix \(X \in \mathbb{R}^{TL\times D_M}\), is first projected to \(H\) heads of \(Q, K, V \in \mathbb{R}^{TL \times D_M/H} \) through a linear layer, where \(D_M\) is the hidden dimension size. These matrices are then fed into the attention mechanism, which essentially computes \(Softmax(QK^T)V\) for each head, resulting in \(H\) matrices of \(O \in \mathbb{R}^{T \times D_M/H}\). Finally, these \(O\) matrices are all-reduced along \(D_M\) dimension and then passed through an FFN layer to generate the block output.
The model parameters primarily come from the QKV projection and the linear layer parameters in the FFN. The immense scale of model parameters (e.g., 3B, 8B, or even over 100B) imposes a significant burden on edge devices, which typically have limited computing power and memory capacity.

Weight quantization (e.g., BF16\cite{micikevicius2017mixed}, INT8\cite{xiao2023smoothquant}, FP8\cite{micikevicius2022fp8}, NvFP4\cite{chmiel2025fp4}, etc.) has been proposed to address the challenges posed by the large model size.
Among these, ternary quantization introduced by BitNet\cite{wang2023bitnet} compresses model weights into a ternary format (a.k.a 1.58 bit) through quantization-aware training(QAT). 
As illustrated in Figure~\ref{fig:bg-attention}(a,b), BitNet replaces all weights in linear modules with ternary values, reducing the model's parameter size by at least eight times compared to the BF16 precision while largely preserving inference accuracy. 

Furthermore, since multiplying by ternary can technically be implemented without multipliers, ternary quantization provides substantial theoretical improvements in computational efficiency. As shown in Figure \ref{fig:bg-mem_prof}(b), compared to A8W8 quantization, the A8W1.58 quantization reduces the computation in the FFN and QKV projection layers by directly eliminating the multiplications.




\subsection{Attention Mechanism Optimization}
Although linear layers can be aggressively quantized, attention layers cannot be quantized in the same way due to the widespread outliers in activations \cite{hooper2024kvquant, xiao2023smoothquant, lin2024awq}.
Moreover, the original Multi-Head Attention (MHA) mechanism suffers from high computational complexity especially in the long-context scenario, due to the quadratic scaling with respect to the input token length, making it unsuitable for edge devices. 

Extensive research has been conducted to alleviate this issue.
Sparse Attention(SA) mechanisms improve the traditional attention mechanism during both the decode and prefilling stages by introducing sparsity by pre-selecting the important QK pairs. For example, the Big Bird\cite{bigbird}, StreamingLLM\cite{streamingllm} and other previous work\cite{li2025radial, yuan2025nsa, guo2024blocksparse, beltagy2020longformer, child2019generating} indicate that the local QK pair(diagonal pattern) and attention sink (vertical pair) dominate the significant QK pairs.
Sparse attention reduces the computational complexity of attention mechanism to linear complexity in terms of the token length by selecting a fixed size of QK pairs per row.
Furthermore, the linear attention mechanism\cite{katharopoulos2020lineartransformers} offers a more radical solution by replacing the traditional attention mechanism with a function that has linear complexity.
Although naive linear attention has limited performance, it remains an active area of research, with numerous ongoing studies demonstrating promising improvements to its architecture \cite{gu2023mamba,peng2023rwkv,mamba2,yang2023gla}.






\subsection{Motivation}\label{sec:motivation}

We analyze the challenges in enhancing the efficiency of real-time Ternary LLM inference and the limitations of existing approaches, which motivate the design of \nickname, encompassing compute core, memory, and system aspects.

\paragraph{Existing ternary GEMM engines are inefficient}
To overcome the inefficiency of deqauntize-based methods, lookup table (LUT)-based approach is proposed to improve the area and power efficiency of low-bit mix-precision matrix multiplication (mpGEMM). 
Substantial work\cite{mo2024lut,park2022lutgemm} has proposed LUT-based PE for mpGEMM with various precision, most of which are tailored for integer format using bitwise encoding methods, which require extensive adders to sum up the partial results of each bit. Some studies\cite{tellme,wang2025bitnet} have also introduced element-wise encoding schemes specifically for ternary mpGEMMs, which often come with the drawback of large table sizes. Our analysis reveals that these methods still fail to fully exploit the symmetric and inherent sparsity of ternary mpGEMM, motivating us to build a more optimal design that achieves balanced trade-offs between adders and lookup table logic (Section \ref{sec:stl}). 


\begin{figure}[!t]
    \centering
    \includegraphics[width=8.4cm]{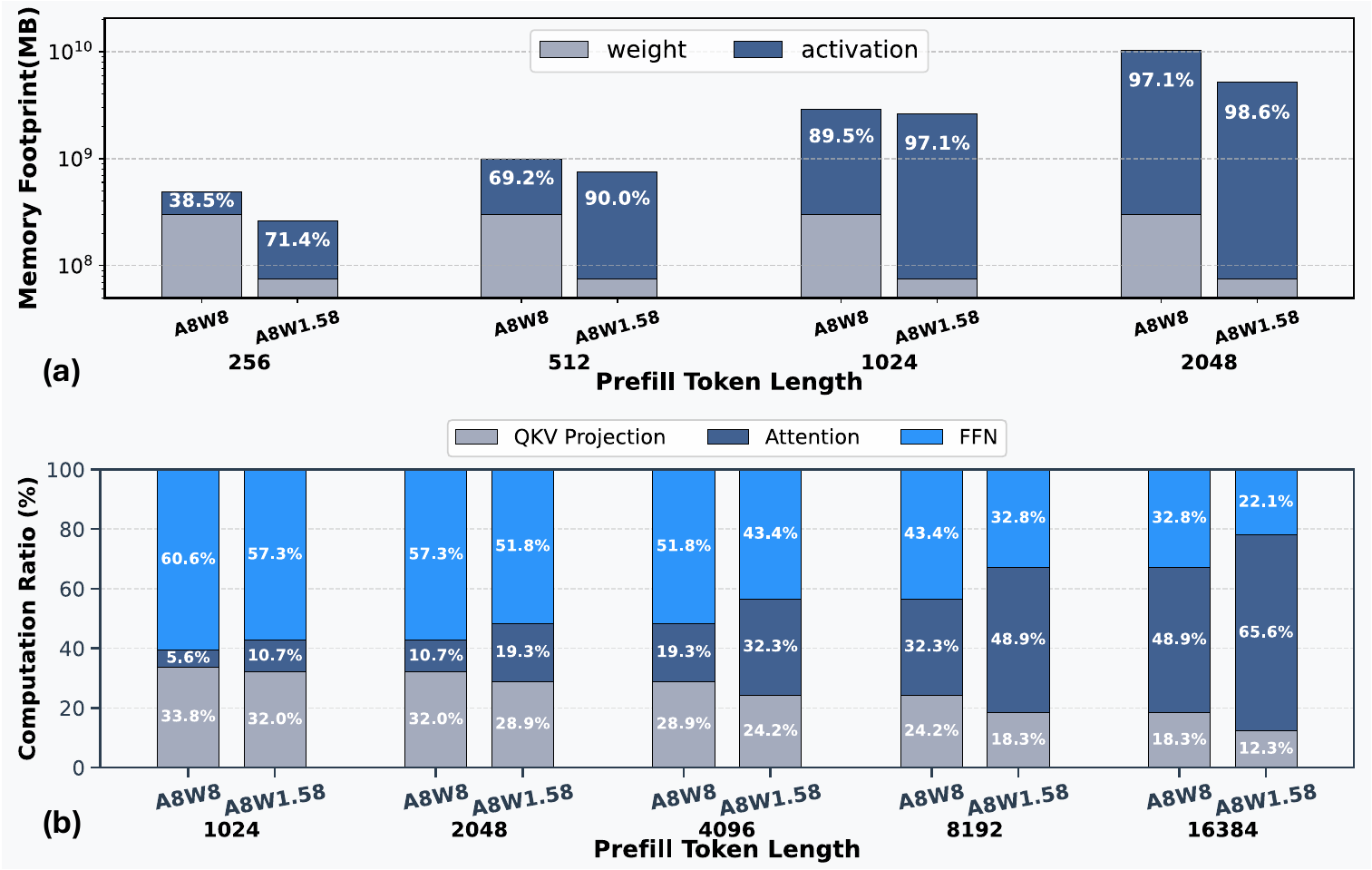}
    \vspace{-10pt}
    \caption{(a) Memory access breakdown of QKV Projection \& Attention with different quantization and prefilling token length in Llama-7B,
    (b) Computation breakdown of Llama-7B with different quantization, each MAC operation in A8W8 consists of one multiplication and one addition.}
    \label{fig:bg-mem_prof}
\end{figure}


\paragraph{Reducing DRAM access is crucial to performance and energy efficiency of low-batch LLM inference}
Although the prefilling stage is well-known as being compute-bound, existing work\cite{FACT,ELSA} often overlook the performance overhead and excessive energy incurred by writing/reading intermediate results to DRAM.
In low-batch inference scenarios, all operators in the attention mechanism exhibit low arithmetic intensity (OP/Byte). For example, during the prefilling phase of the Llama-7B model with a $TL$ of 1024, the arithmetic intensity of the QK matmul (114) is 10 times lower than that of the Q/K/V projection (1024)\cite{llmunveiled}, resulting in a memory bound issue on the A6000 GPU. 
Additionally, accessing DRAM incurs an energy cost over 300$\times$ higher than computation per data\cite{tpuv4i}. 
As shown in Figure \ref{fig:bg-mem_prof}, assuming that $TL$ is 1k in Llama-7B with ternary quantization, loading/storing the activation matrices (including Q/K/V, attention score, etc.) incurs 97\% of total memory access in each attention head with corresponding QKV projection, leading to substantial energy consumption.

In the decoding stage, loading weights becomes the main bottleneck due to the auto-regression token generation. As a ternary value can technically be represented with 1.58 bits, existing work fail to achieve an optimal packing density.

In summary, to maximize the performance and energy efficiency of ternary LLMs, it is essential to reduce the memory accesses of both static weights and dynamic activations, which will be introduced in Section \ref{sec:stl-twd} and Section \ref{sec:sys-lf} respectively.

\paragraph{Attention mechanism poses challenges to ternary LLM inference system efficiency}
The quadratic complexity of the attention mechanism also poses computational challenges. Since low-precision quantization is difficult to apply to attention, high-precision GEMM cores are still required. As shown in Figure \ref{fig:bg-mem_prof}(b), during the prefilling stage, attention computation becomes increasingly dominant with longer token sequences, leading to workload imbalance between low- and high-precision cores.
Prior work featuring heterogeneous GEMM cores \cite{tellme,moLUTTensorCore2025b} overlook the overall core utilization, thereby limiting system efficiency.
We observe that leveraging sparse attention enables a balanced computational workload between the QKV linear projection and the attention mechanism.
To exploit this, we design a heterogeneous architecture with cross-stage co-optimized dataflow between QKV projection and attention, as detailed in Section \ref{sec:sys}.

\section{Compute and Memory Optimization for Ternary Linear Layer}\label{sec:stl}

\begin{table}[!t]
    \centering
    \begin{threeparttable}
    \caption{Ternary Compute Core Complexity Comparison \textbf{($S_a<1$)} } 
    \renewcommand{\arraystretch}{1.5}  
    \small  
    \begin{tabular}{@{\hspace{5pt}}l@{\hspace{5pt}}c@{\hspace{5pt}}c@{\hspace{5pt}}c@{\hspace{5pt}}}
        \toprule
        \textbf{Component} & \textbf{\makecell{Pre-\\Compute}} & \textbf{\makecell{Look\\Up Logic}} & \textbf{\makecell{Adder\\Tree}} \\ \midrule
        \textbf{Add Only} & / & / & $O(N_tGg)$ \\
        \textbf{General LUT}  & $O(\frac{G2^gg}{N_t})$ & $O(2N_tG2^g)$ & $O(N_t(G+g))$\\
        \textbf{Ternary LUT} & $O(\frac{G3^gg}{N_t})$ & $O(N_tG3^g)$ & $O(N_tG)$\\
        \textbf{STL (Ours)} & $O(\frac{S_aG2^gg}{N_t})$ & $O(S_aN_tG2^g)$ & $O(S_aN_tG)$\\
        \bottomrule
    \end{tabular}
    \vspace{-10pt}
    \label{table:core_compare}
    \end{threeparttable}
\end{table}



\begin{figure*}[!th]
    \centering
    \includegraphics[width=18cm]{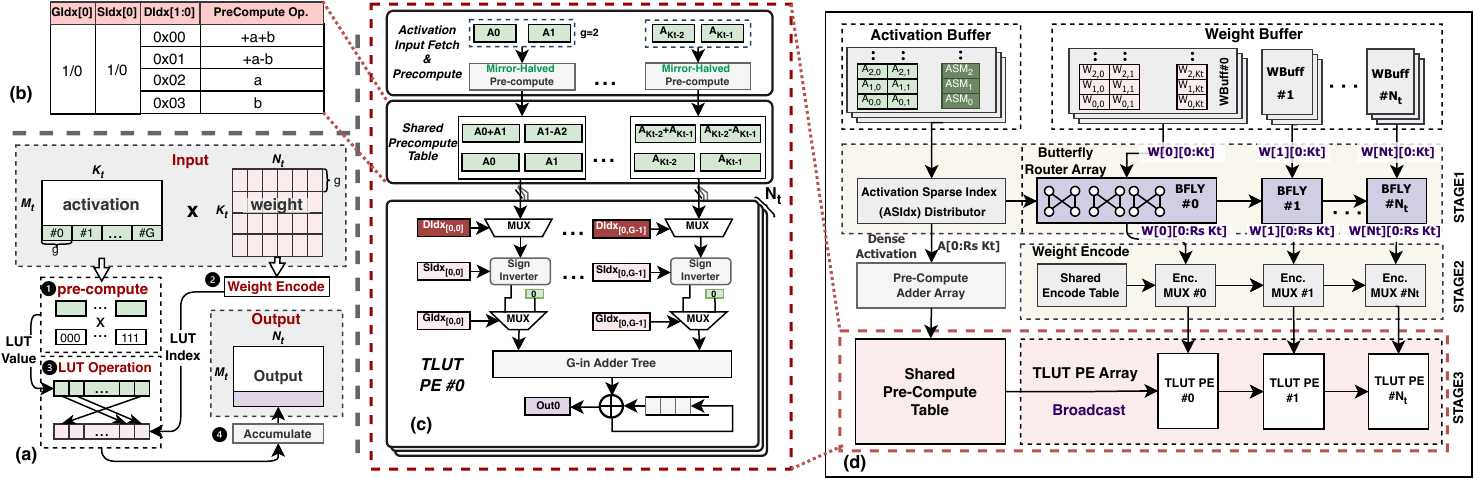}
    \vspace{-10pt}
    \caption{
    (a) LUT-based compute flow of a mpGEMM tile with input activation with shape of ($M_t,K_t$) and weight with shape of ($K_t,N_t$), and generating the result with shape of ($M_t,N_t$). 
    (b) The zero-aware ternary pre-compute table encoding with input grouped activation data \{a,b\}. 
    (c) The TLUT PE architecture where $N_t$ TLUT PEs share the precompute logic and table.
    (d) The overall 3-stage pipeline architecture of STL Core with input buffers. }
    \label{fig:stl-all}
\end{figure*}

\begin{figure}[!t]
    \centering
    \includegraphics[width=7.8cm]{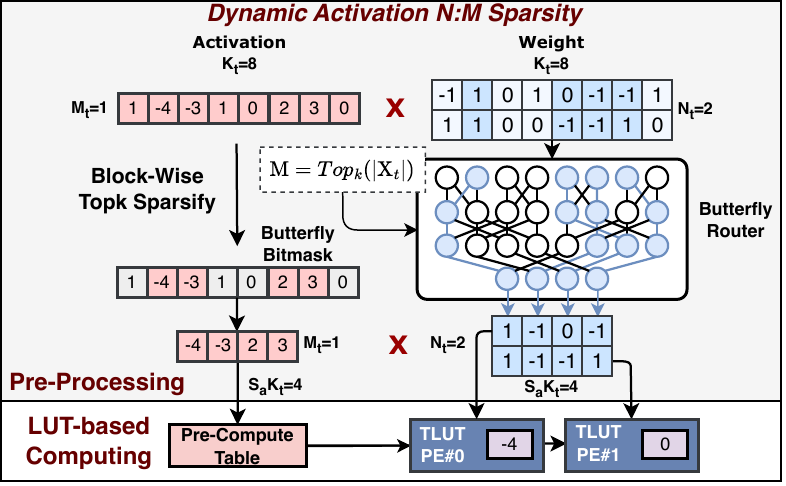}
    \vspace{-5pt}
    \caption{Compute flow of Dynamic Activation N:M Sparsity in STL Core }
    \label{fig:sparsity}
\end{figure}

\subsection{LUT-Based Ternary mpGEMM Analysis}\label{sec:stl-compare}

Ternary mixed-precision matrix multiplication can be realized through various hardware architectures.
Given a high-precision input activation of shape (M$_t$, K$_t$) and a ternary weight matrix of shape (K$_t$, N$_t$), the resulting output matrix has shape (M$_t$, N$_t$), where M$_t$, N$_t$, and K$_t$ represent the core's parallelism along three dimensions.
Considering a simple dot product between vector A(1, K$_t$) and vector W(K$_t$, N$_t$), the most basic implementation, referred as the add-only core, flips the sign bit of the input activation according to the ternary weight's most significant bit (MSB) and then adds them together through an adder tree. However, this approach incurs substantial overhead from the adder tree.
Alternatively, LUT-based methods reduce adder tree complexity by partitioning the vector into chunks and computing each chunk's dot product using precomputed lookup tables. As shown in Figure \ref{fig:stl-all}(a), 
in LUT-based computing, the activation and weight tiles are first split along the $K$ dimension into small groups of size $g$, resulting in $G=K_t/g$ groups. Each activation group then performs the following steps: \blackding{1} all possible dot products between the group’s activations and bit-wise weights are precomputed and stored in a table;  \blackding{2} the weight matrix is bit-serialized to generate LUT indices; \blackding{3} partial dot product results are retrieved from the LUT with the indices;  \blackding{4} The partial results are accumulated to produce the final result.

A general LUT method encodes each ternary weight group in INT2 format, decomposes it into two INT1 groups, and computes for each group separately with lookup tables\cite{mo2024lut}. However, this approach fails to exploit the element-wise symmetry of ternary data, as it uses four states to encode only three valid values, introducing unnecessary complexity in the lookup logic and adder tree.
Prior work \cite{tellme} proposes a ternary-specific LUT core using a base-3 table to store all possible dot product results with the ternary weight group. However, this method incurs large lookup table overhead.
For a $g$ of 3, the base-3 table size can reach up to 27. Even with the mirror consolidation technique \cite{wei2025t}, which folds the table based on the sign bit, the table size remains 14.

In summary, as shown in Table \ref{table:core_compare}, the primary overhead of the LUT-based computing logic arises from three components: pre-computation logic, table lookup logic, and adder logic.
The existing LUT-based approaches fail to achieve an optimal balance between lookup table overhead and adder tree complexity.
To address this limitation, we introduce the Sparse Ternary LUT (STL) core, which comprises an array of Ternary LUT (TLUT) Processing Elements (PEs) and a structure to support Dynamic Activation N:M Sparsity (DAS).

\subsection{Ternary LUT Processing Element}\label{sec:stl-pe}

Figure \ref{fig:stl-all}(c) illustrates the architecture of the proposed TLUT Processing Element. To fully exploit the symmetry of ternary weights, we adopt an element-wise approach rather than a bitwise one. 
We begin by proposing a zero-aware ternary precompute table, as shown in Figure \ref{fig:stl-all}(b).
To exploit the prevalence of zero-valued blocks in ternary weight matrices, we augment the encoding table with a 1-bit sparse gate index ({\em GIdx}). GIdx is asserted when an entire weight group is zero, suppressing all downstream computation and eliminating the associated dynamic power. 
Precomputed results corresponding to non-zero (dense) weight blocks are handled by a 2-bit dense index ({\em DIdx}) that selects one of four pre-computed, symmetric partial products, while a 1-bit sign index ({\em SIdx}) flips the selected value when the negative mirror is required. The concatenated 3-bit tuple {SIdx, DIdx} thus encodes the complete set of eight possible dense outputs.

As shown in Figure \ref{fig:stl-all}(c), the ternary weights are first decoded into DIdx, GIdx, and SIdx, while the input activations are divided into $G=K_t/2$ groups. 
These groups are then fetched to the mirror-half pre-compute adder logic, which constructs the shared pre-compute table.
Within each TLUT PE, DIdx steers a 4-to-1 MUX, SIdx conditions a sign inverter, and GIdx gates a 2-to-1 MUX, together forming the pipeline within the PE. The $G$ partial results are then fed into the adder tree for accumulation, ultimately generating the tile output $Y$.
To amortize the expensive adder logic for pre-computation and register arrays for the tables, the table output ports are broadcast to $N_t$ TLUT PEs, thereby enabling parallel output $Y[0,N_t]$ and table reuse in the linear layer of the ternary LLM.

\subsection{Dynamic Activation N:M Sparsity} 

In LLMs, the hidden dimension ($K_t$) is typically enormous (1K-10K), especially in the FFN layer.
Intra-token sparsity can be exploited to reduce the computation complexity as well as LUT overhead.
Since a ternary weight takes only three possible values, ~30\%-40\% of the weight value are zeros. However, their completely irregular distribution makes it impossible to leverage them efficiently.



Alternatively, as illustrated in Figure~\ref{fig:sparsity}, we exploit the sparsity within token activation to optimize the lookup table size. The input activations are divided along the $K$-dimensional into multiple blocks of size $B_s$. Within each block, we construct a N:M sparsity pattern based on TopK, with the sparse ratio $S_a$ defining the proportion of valid activation. The TopK bitmask serves as an index to select weight channels, thereby reducing the number of groups $G$ in LUT computation, as shown in Table \ref{table:core_compare}.
From the algorithm perspective, we finetune the pre-trained model with sparsity mask, which can be expressed as a sparsify-then-quantize function:
\begin{equation}
    Y = (\text{Q}_{\text{int8}}(X) \odot M \cdot \text{Q}_{\text{1.58}}(W)^{\top}, M=\text{Top}_k(|X|)
\end{equation}
The TopK operation is performed along the hidden dimension of the activation in block units of size $B_s$. $\text{Q}_{\text{1.58}}$ and  $\text{Q}_{\text{int8}}$ are ternary and INT8 quantization functions for weights and activations. 
The N:M sparse bitmask $M$ identifies valid activations within the block based on the TopK results computed over the absolute values of the block activations.

In the hardware implementation, weight channels corresponding to valid activations are dynamically extracted using a butterfly router. The router's control signals are generated using the TopK sparse bitmask {\em ASM} of the activations, retrieved from the activation buffer, as shown in Figure \ref{fig:stl-all}(d). The complexity of the butterfly router is $O(k \log k)$ with $k$-input, which is lower than that of commonly used compression networks such as BENES ($O(2k \log k)$) or crossbar ($O(k^2)$)\cite{sigma2020hpca}.




\begin{figure}[!t]
    \centering
    \includegraphics[width=8.8cm]{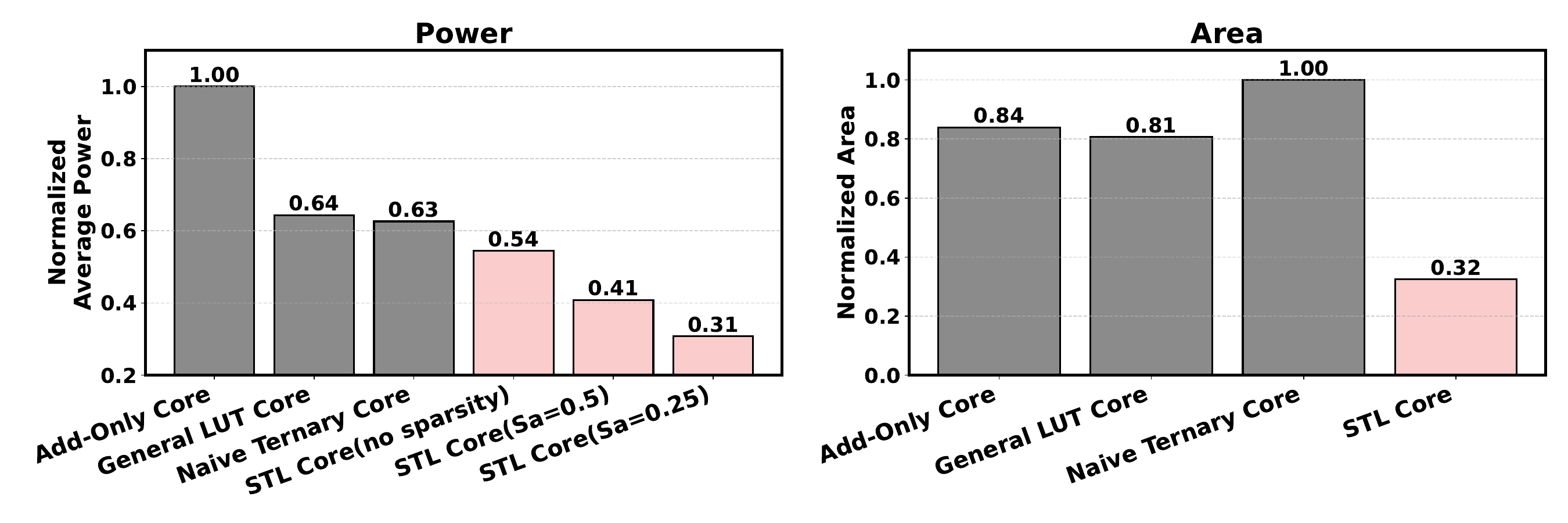}
    \vspace{-20pt}
    \caption{Area and power overhead of different A8W1.58 core}
    \label{fig:core_comp}
    \vspace{-10pt}
\end{figure}

\subsection{Sparse Ternary LUT Core}\label{sec:stl-core}





Figure \ref{fig:stl-all}(d) shows the STL Core architecture. Input activations and ternary weights (packed in 2-bit format) are stored in the input and weight buffers, respectively. 
In each cycle, the STL Core gets an activation of shape [1, $K_t$] and weights of shape [$K_t$, $N_t$] from the buffers. These inputs are then processed through a three-stage pipeline: 

\textbf{Stage 1:}
The dense activation with ASM and the complete weight matrix are read from the buffer. The block-wise ASM is transformed to the control bits for the butterfly array to dispatch the weights corresponding to non-zero activations. 
\textbf{Stage 2:}
The dense activation then passes through precomputed logic to generate lookup table data, which is fetched into the shared precompute table. Meanwhile, for each of the $N_t$ TLUT PEs, DIdx, SIdx, and GIdx are extracted from the corresponding weights.
\textbf{Stage 3:}
The precompute table and TLUT PEs operate collaboratively to generate the final tile output.
The final result is accumulated within each PE to accommodate hidden dimensions of arbitrary length. Then it is converted to FP16 format for subsequent vector-wise operations or quantization.

The STL core is optimized for GEMV operation by computing one activation vector at a time. During the decoding stage of one-batch LLM inference, the majority of tensor operations are GEMV. Compared to conventional tensor core designs optimized for GEMM\cite{nvidia}, our design choice ensures higher utilization of the compute hardware. 

\begin{figure}[!t]
    \centering
    \includegraphics[width=8.8cm]{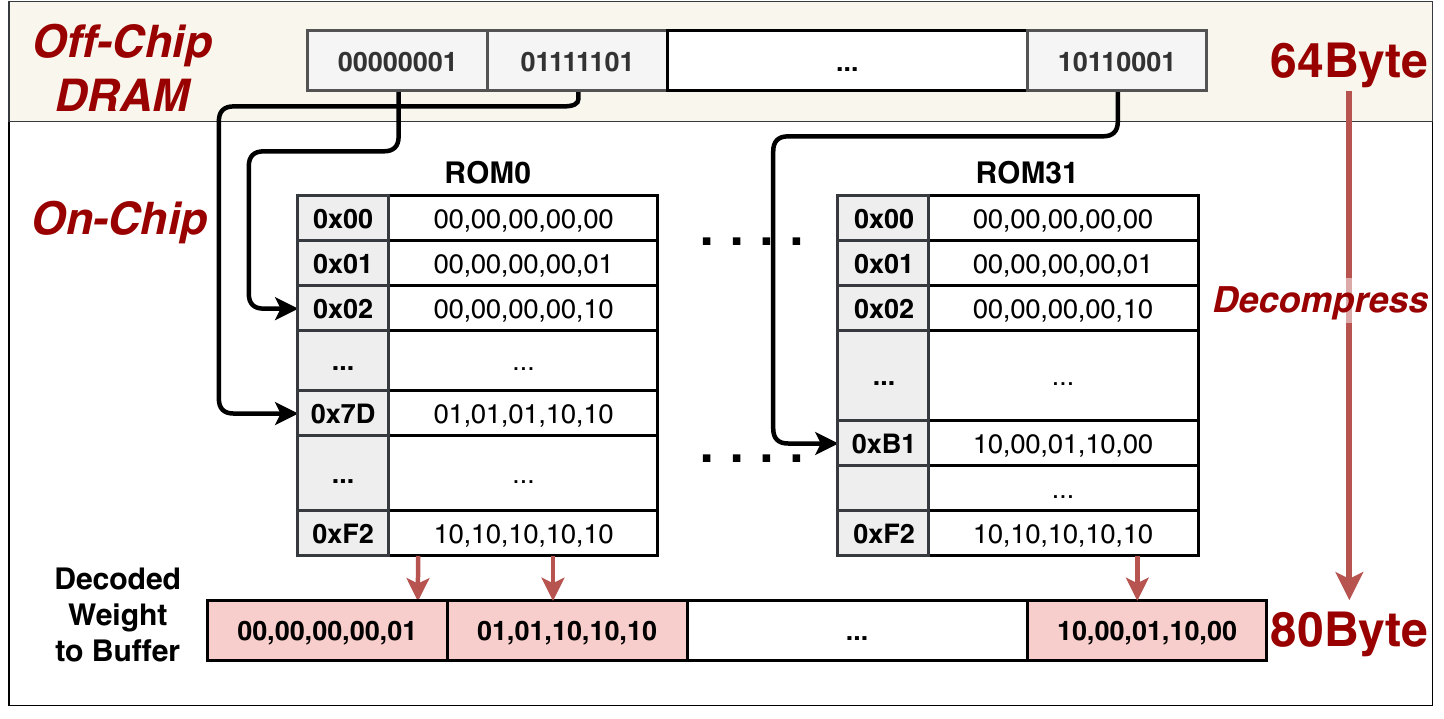}
    \vspace{-20pt}
    \caption{LUT-based 64B:80B Ternary Weight Decompression}
    \label{fig:stl_tmc}
\end{figure}

Figure~\ref{fig:core_comp} compares the power and area overhead of four core architectures, the add-only core, general LUT tensor core\cite{moLUTTensorCore2025b}, the naive ternary LUT core\cite{tellme} used in prior work, and our STL core, configured with identical compute throughput.
For the STL Core, we report power consumption under three different sparsity levels: no sparsity, 50\% sparsity ($S_a=1/2$) and 75\% sparsity ($S_a=1/4$). 
All cores operate at 500MHz and are synthesized using 28nm technology.

Although the add-only core eliminates all multipliers in dot product, the adder tree with a fan-in size of $K_t$ is still expensive. 
In terms of power consumption, both the LUT Tensor Core and Naive Ternary Core reduce the power overhead of the large adder tree through LUT logic. However, due to their large precompute table sizes, they do not show significant area benefits.
The element-wise zero-aware encoding enables the STL core to minimize the table size while optimizing the adder tree's fan-in. The proposed STL core achieves a 52\% reduction in area and a 46\% reduction in power consumption without sparsity compared to the add-only core.
Enabling sparsity further reduces the power consumption by lowering the dynamic power of TLUT PEs, while the butterfly router introduces minimal hardware overhead.

\subsection{LUT-based 60B:80B Ternary Weight Decompression}\label{sec:stl-twd}



To eliminate redundant memory access caused by the standard 2-bit data packing method\cite{t-mac, wang2024ladder}, \nickname incorporates a dedicated Ternary Weight Decompression (TWD) module. 
Due to the fact that each ternary element carries only 1.58 bits of information which is calculated by log(3)/log(2), we can use an 8-bit binary index ($2^8=256$) to express five ternary weights ($3^5=243<256$), which effectively express each weight in 1.6 bits.
Weights are encoded offline and stored in DRAM. As illustrated in Figure \ref{fig:stl_tmc}, the TWD module is integrated into the memory interface, where it performs decompression during the weight pre-fetching stage. This process leverages efficient lookup logic implemented using multiple dual-port ROMs, without causing additional delay\cite{wang2024ladder}.
Being byte aligned, our compression method is naturally aligned with the ternary LLM training method, without causing accuracy issue due to the inconsistency between the alignment method and the original method\cite{wang2025bitnet}.

Although this optimization only appears to reduce one-fifth of the off-chip weight memory access compared to a standard 2-bit encoding, this reduction directly reflects on end-to-end latency in the decoding stage due to its memory-bound nature.

\section{System-Level Optimization for Ternary LLM}\label{sec:sys}

\subsection{\nickname Heterogeneous Architecture Overview}\label{sec:sys-all}

\begin{figure}[!t]
    \centering
    \includegraphics[width=7.8cm]{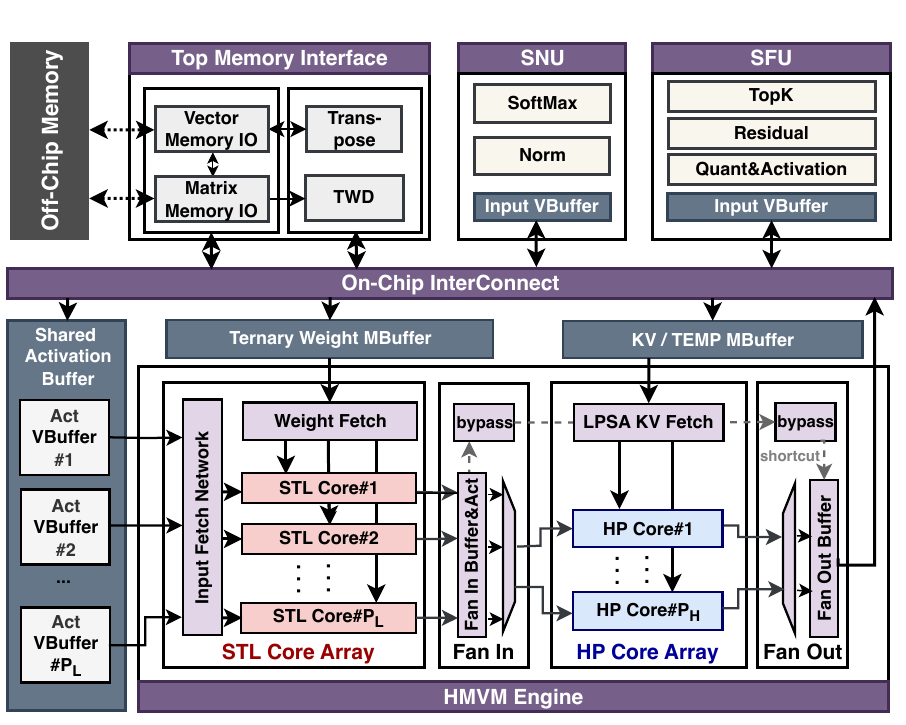}
    \vspace{-10pt}
    \caption{\nickname accelerator system architecture}
    \vspace{-10pt}
    \label{fig:arch}
\end{figure}
Figure \ref{fig:arch} shows an overview of \nickname sparsity-aware heterogeneous architecture.
The Heterogeneous Matrix Vector Multiplication (HMVM) engine handles all GEMM computations. It comprises an STL Core Array and a High-Precision (HP) Core Array, both sharing a buffer for input vectors. Each core array is equipped with a dedicated weight buffer for ternary weight and temp data buffering (including Key and Value tensors in the attention mechanism).
The STL Core Array, central to the \nickname architecture, operates with a parallelism factor $P_{L}$, while $P_{H}$ HP Cores support high-precision matrix computations such as $QK^{\top}$ and $SV$ in attention mechanisms. 
Both core arrays can operate independently or collaborate in a pipelined manner, enabled by the Fan-In/Out interconnect and bypass datapath.
The Special Function Unit (SFU) is responsible for activation, residuals, and sparse Top-K operations in DAS, and is directly connected to the HMVM Engine output. The Softmax \& Normalization Unit (SNU) handles vector-wise softmax and normalization operations.
At the top of the \nickname processor, the Top Memory Interface (TMI) orchestrates memory transfers for matrices (e.g., weights, KV) and vectors (e.g., activations) between on-chip and off-chip buffers. It manages both the Vector Buffers (VBuffer) and Matrix Buffers (MBuffer), with Transpose and TWD modules integrated.

\nickname adopts an algorithm-hardware co-design strategy to enhance on-chip throughput and reduce memory accesses, thereby harnessing the energy efficiency benefits of sparse ternary LLMs. First, we identify data reuse opportunities enabled by sparse attention and propose the Linear-Projection-Aware Sparse Attention (LPSA) dataflow for cross-stage co-optimization (Section \ref{sec:sys-lf}). Second, we detail the design of the configurable HMVM engine (Section \ref{sec:sys-collab}). Finally, we develop a design space exploration (DSE) method for the \nickname architecture, enabling comprehensive analysis across both algorithmic and hardware dimensions (Section \ref{sec:sys-dse}).

\subsection{Linear-Projection-Aware Sparse Attention Dataflow}\label{sec:sys-lf}

\begin{figure}[!t]
    \centering
    \includegraphics[width=8.8cm]{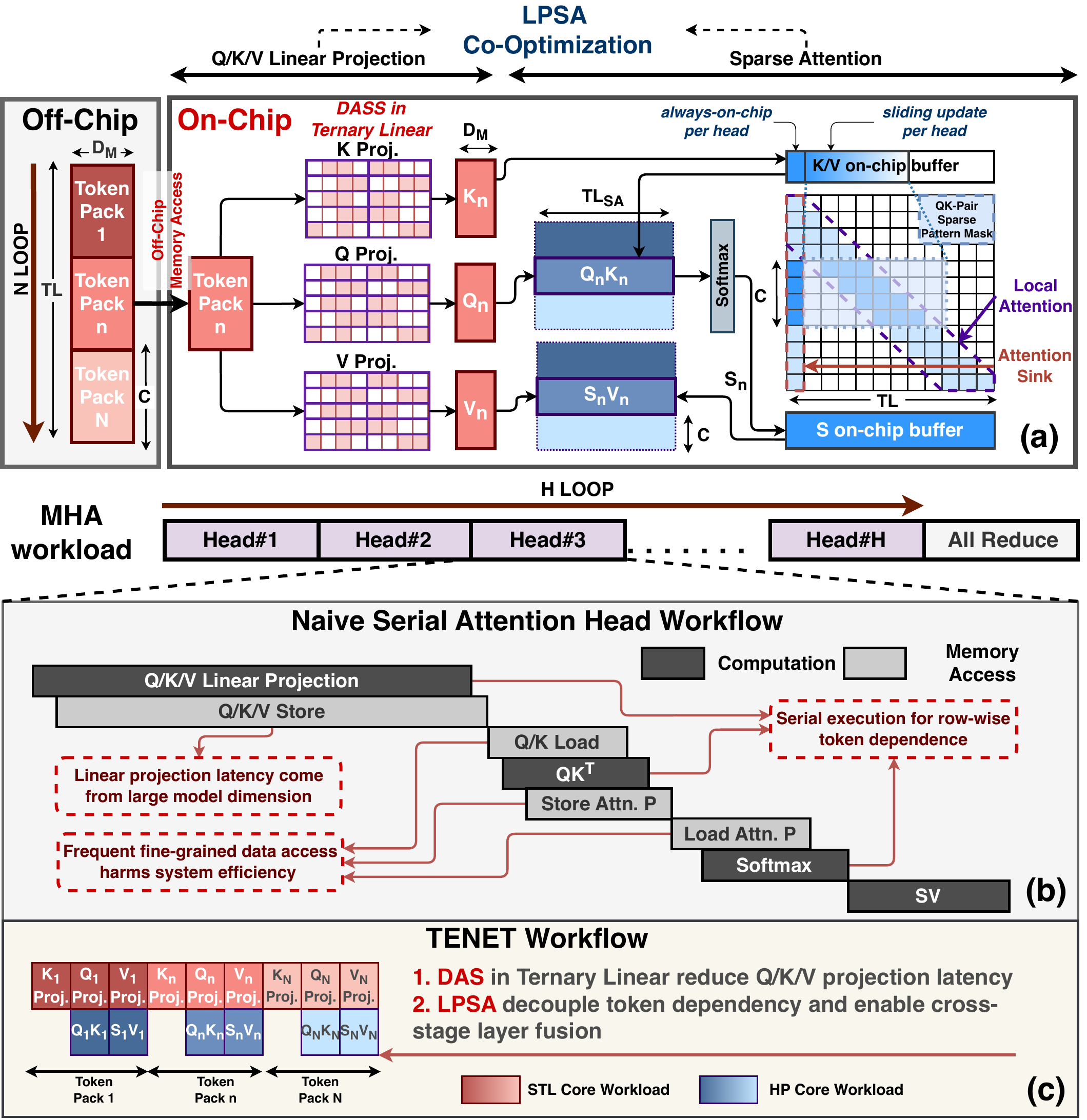}
    \vspace{-10pt}
    \caption{
    (a) The LPSA workflow within an attention head.
    (b) The naive serial attention head work flow suffering from high latency and excessive memory access of intermediate data.
    (c) The \nickname workflow combining DAS and LPSA, forming a fine-grained pipeline which optimizes both memory access and hardware utilization.
    }

    \label{fig:sys-lpsa}
    \vspace{-10pt}
\end{figure}

The naive ternary LLMs\cite{wang2023bitnet, maBitNetB1582B4T2025} neglect the optimization of the attention mechanism. As the token length increases, the attention layer dominates both memory traffic and FLOPs.
As shown in Figure \ref{fig:sys-lpsa}(b), the naive workflow executes the layers sequentially and stores excessive intermediate activations from the attention layer to the off-chip memory due to the limited SRAM capacity. This incurs extra latency and limits the overall utilization of STL Cores and HP Cores.

To address this, we propose Linear-Projection-Aware Sparse Attention (LPSA) dataflow to fuse ternary linear projection with sparse attention at tile granularity.  
Algorithm~\ref{alg:sa-hslf} summarizes the prefilling phase of multi-head attention under LPSA.
Following prior sparse-attention templates~\cite{bigbird,streamingllm,beltagy2020longformer}, we adopt the SA pattern shown on the right of Figure~\ref{fig:sys-lpsa}(a) to balance both algorithmic and hardware efficiency: each head computes a window of diagonal QK pairs (local attention) and a column of QK pairs at the sequence start (attention-sink).
Only the KV pairs of the small fixed attention sink and a fixed width of local attention are retained on-chip. 
We use an SA mask, $M_{k,v}$, to indicate $TL_{SA}$ valid KV pairs per token for each pack.
We partition the attention computation of $TL$ tokens into $N$ independent token packs of length $C$. 
Within a pack, QKV projection and attention consume (i) the incoming C tokens from DRAM, (ii) the on-chip KV sink and window and (iii) the preloaded QKV weights for one head.

\begin{algorithm}[!b]
\begin{small}
  \caption{Linear-Projection-Aware Sparse Attention Dataflow Forward Pass with DAS during Prefilling Stage}
  \label{alg:sa-hslf}
  \begin{algorithmic}[1]
    \Require{Hidden state: $X\in \mathbb{R}^{TL\times D_mS_a}$, projection weight: $W_q[1:H],W_k[1:H],W_v[1:H]\in \mathbb{R}^{D_m\times D_h}$, head number $H$, local chunk size $C\in[TL]$, KV sparse mask $M_k,M_v\in\{0,1\}^{N\times C\times TL}$}, activation sparse mask $ASM\in\{0,1\}^{TL\times D_m}$
    
    \State Divide $X$ and $ASM$ into $N=\frac{L}{C}$ blocks ${\{X{[1]},\ldots,X{[N]}} \}\in \mathbb{R}^{C\times D_mS_a}$, ${\{ASM{[1]},\ldots,ASM{[N]}} \}\in \{1,0\}^{C\times D_m}$
    \State On-Chip initialize non-local temp KV buffer $K_{Buf},V_{Buf}\in \mathbb{R}^{(TL_{SA}+C)\times D_h}$
    \For{$h \leftarrow 1, H$}
        \State Load $W_q{[h]}, W_k{[h]}, W_v{[h]}$ from DRAM to SRAM
        \For{$n\leftarrow 1,N$}
            \State Load $X[n],ASM[n]$,$M_{k,v}[n]$ from DRAM to SRAM
            
            \State $K{[n]} \gets {X[n](W_k{[h]\odot ASM[n])}}$ \Comment{STL Core}
            
            \State $K_{Buf} \gets \{{K_{Buf}, K[n]}\}\odot M_k[n]$ \Comment{Update K buffer}
            \State $Q{[n]} \gets X[n](W_q{[h]}\odot ASM[n])$ \Comment{STL Core}
            \State $P{[n]} \gets Q{[n]}(K_{Buf}{^{\top}\odot M_k{[n]})}$ \Comment{HP Core}
            \State $S[n]  \gets \text{SoftMax}(P{[n]})$ \Comment{SNU, stored to S Buffer}
            \State $V[n] \gets X[n](W_v{[h]}\odot ASM[n])$ \Comment{STL Core}
            \State $O^h[n] \gets S{[n]}(\{V{[n],V_{Buf}\}\odot M_{v}[n])}$ \Comment{HP Core} 
            \State $V_{Buf}  \gets \{{V_{Buf}, V[n]}\}\odot M_v[n]$ \Comment{Update V buffer}

            \State Store $O^h[n]$ to DRAM
        \EndFor
        
      \State \Return $O^h=\{O_{[1]}, \ldots,O_{[N]} \}$
    \EndFor
    \State \Return $O=\{O^1, \ldots,O^H \}$
  \end{algorithmic}
\end{small}

\end{algorithm}

Algorithm~\ref{alg:sa-hslf} (lines 4–15) and Figure~\ref{fig:sys-lpsa}(a) detail the per-head execution: First, the ternary QKV weights are loaded once and remain resident. 
Once a token pack and its associated ASM and SA mask are loaded to on-chip SRAM, the $K$ matrix is produced and inserted into the KV buffer.
Then the corresponding Q vector is generated and immediately consumed by the sparse $QK^\top$. 
A similar flow is used to fuse the computation of $V$ projection and $SV$. To resolve the data dependency associated with left multiplication, $SV$ is transformed into the equivalent $V^\top S^\top$.
During decoding, the flow is similar but simplifies to $N=1$.
Furthermore, LPSA’s cross-layer optimization is well-suited for Linear Attention-based models, as each token's attention size is inherently constant\cite{yang2023gla, katharopoulos2020lineartransformers}.
Incorporating DAS for ternary linear layers, as shown in Figure \ref{fig:sys-lpsa}(c), the \nickname architecture reduces the overall latency by (i) reducing the Q/K/V projection latency and (ii) overlapping linear projection and attention computation to conceal attention latency.

\subsection{HMVM Engine and Instruction Sets}\label{sec:sys-collab}

The function of HMVM Engine can be interpreted as cascading a ternary mpGEMV with a high-precision GEMV. Specifically, for an input vector $x$, the output is computed as \(y=(xA_{ternary})B_{hp}\). In \textbf{Standard Mode}, when a sequence of input vectors is stacked into a row-major matrix $X$, the HMVM Engine iterates through the rows to perform cascaded GEMMs, resulting in \(Y=(XA_{ternary})B_{hp}\). However, it is common for one of the matrices to require transposition before multiplication. 
For example, in the MHA mechanism, according to the LPSA workflow, we need to calculate both $QK^\top =(XW_{Q})K^\top$ and $V^\top S^\top=(XW_{V})^\top S^\top$ for left-multiplication data dependency. Notably, the intermediate result $V$ requires transposition, whereas $Q$ does not.

A conventional way to deal with matrix transposition is to store the intermediate result back to the memory in a transposed format and then read it back for the second GEMM, resulting in low hardware efficiency.
To address this issue, we introduce \textbf{Transposed Mode}. Instead of performing the first GEMM and then transposing its output matrix, we allow the STL Core to compute the first GEMM column by column by properly supplying the operands from the buffers. In this way, the output of the first GEMM is naturally transposed before feeding to the HP Core.

To enable full programmability, we introduce the FMvMul instruction as an extension of the existing MvMul ISA \cite{brainwave}: \\
\begin{normalsize}
    \setlength{\baselineskip}{10pt} 
    \[\textbf{FMvMul.\{En$_{STL}$, WeightSAddr\}\{En$_{HP}$,KvSAddr\}}\]
\end{normalsize}
    
En$_{STL}$ and En$_{HP}$ are two flag bits that indicate whether the STL Core and HP Core are enabled, respectively. One core is bypassed if its corresponding flag is disabled.
During each iteration along the M$_t$ dimension, the parameters WeightSAddr and KvSAddr specify the read start addresses for the Ternary Weight Buffer and KV/TEMP Buffer. The iteration counts for the K$_t$ and N$_t$ dimensions, as well as the selection between Standard Mode and Transposed Mode, are configured by writing values to specific registers using the $s\_wr$ instruction.

%



\subsection{Algorithm-Hardware Design Space Exploration}\label{sec:sys-dse}

In the LPSA dataflow, the window size of the selected QK pairs, the parallelism of STL Core and HP Core form an interesting design space.
A larger window size brings lower perplexity but requires more HP Cores to maintain throughput, as well as larger KV Buffers to store intermediate activations, increasing hardware overhead.


To simplify the design, each STL Core and HP Core are configured with the same throughput, denoted as $Thpt$. Assuming that we have $P_L$ STL Cores and $P_H$ HP Cores running in parallel, the latency of the STL Core and HP Core for computing the Q/K/V projection and QK/SV operators can be described as follows:


\begingroup
\small

\begin{align} \label{equ:linear}
    \text{Lat}_{Q\_Proj}/\text{Token} &= (D_h\times Dm)/(Thpt\times P_L) \\
    \text{Lat}_{QK^{\top}}/\text{Token} &= (D_h\times TL_{SA})/(Thpt\times P_H)
\end{align}
\endgroup


To maximize utilization of the \nickname accelerator, we must ensure that the computation latency of the STL Core Array consistently hides the latency of the HP Core Array across various inference scenarios (i.e., $\text{Lat}_{Q\_Proj}/\text{Token} >\text{Lat}_{QK^{\top}}/\text{Token}$). Meanwhile, our goal is to optimize the overall IPJ. 
In summary, the optimization objective function for the entire DSE method can be expressed as follows:


\begingroup
\small
\begin{align}
    &\min \mathcal{L}(C) = PPL \times \ \mathcal{L}_{\text{power}} \times  \mathcal{L}_{\text{latency}} \\
    &\mathcal{L}_{\text{power}}=\sum(P_{STL}(P_L), P_{HP}(P_H), P_{kv\_buf}(TL_{SA}),P_C) \label{alg:l-pe} \\  
    &\mathcal{L}_{\text{latency}}=g(P_L, Param_{LLM}) \label{alg:l-kv} \\ 
    &\text{s.t.} ~    P_L/P_H < D_m/TL_{SA} \label{alg:l-cst}
\end{align}
\endgroup

$C$ is the hyperparameter vector composed of $P_L,P_H,TL_{SA}$. $PPL$ represents the model's perplexity. $ \mathcal{L}_{\text{power}}$ denotes the overall power of the \nickname processor, composed of the power of STL Cores, HP Cores, KV Buffers, and a constant power $P_C$ which includes all other components (Eq. \ref{alg:l-pe}). 
$\mathcal{L}_{\text{latency}}$ denotes the end-to-end latency of generating one token, determined by $P_L$ under the constraint that the computation latency of attention layers are completely hidden (Eq. \ref{alg:l-kv}, \ref{alg:l-cst}).
The optimization algorithm is implemented through grid search, based on the performance model, power model, and the results from the algorithm ablation experiments, which will be discussed in Section\ref{sec:abl}.

\section{Evaluation}

\subsection{Experimental Setup}


\textbf{Algorithm Part:} 
We compare the sparse version of BitNet with the dense BitNet\cite{wang2023bitnet} and our reproduced FP16 LLaMA LLM at two representative model sizes (1.3B and 3B) on edge devices. The sparse BitNet models are continue-trained from dense BitNet using hybrid quantization and activation sparsification. 
Both models are trained with 100B tokens from the RedPajama\cite{weberRedPajamaOpenDataset2024} dataset to ensure a fair comparison.
For DAS, we choose a 50\% N:M sparsity ratio $S_a$ with a block size of 32 to the activations based on the ablation study in Section\ref{sec:abl}. For sparse attention, we set the number of attended tokens per row $TL_{SA}$ during both prefilling and decoding stages to 1024, based on the hardware architecture configuration and ablation study. 

We apply the same optimization to Linear-Attention-based GLA\cite{yang2023gla} model to demonstrate the generality of the TENET architecture. Specifically, we fine-tune the official 340M and 1.3B models following the original training procedure using the Fineweb-Edu dataset\cite{lozhkov2024fineweb-edu}, perform Ternary Quantization(TQ) on the model weights, and apply DAS to the input activations.

\textbf{Hardware Part:} We provide two implementations of TENET on FPGA and ASIC platforms, with the FPGA version serving as a validation prototype for the ASIC design.
For TENET-FPGA, we validate our design on an Intel Stratix 10 MX FPGA and report resource utilization, power efficiency and clock frequency results based on the final post-fit report from Quartus. 
The TENET-FPGA accelerator operates at 400MHz.
TENET-ASIC is designed to fully demonstrate the energy and area efficiency that can be achieved on customized hardware using the TENET architecture.
We synthesize the accelerator core RTL design using Synopsys Design Compiler with a 28nm CMOS technology. 
The core part of TENET-ASIC system operates at 500MHz, with on-chip buffers utilizing foundry-provided SRAM. Power consumption and area results are estimated from the synthesis reports.
We simulate the off-chip memory power with DRAMSim3\cite{dramsim3}.
Additionally, we develop a cycle-accurate simulator and verify it against the implementation of TENET-FPGA prototype. 
Both TENET-ASIC and TENET-FPGA use identical HBM2 memory with ~512 GBps of bandwidth. However, TNET-ASIC has twice the compute throughput due to TNET-FPGA being constrained by the FPGA resources.

For comparison with GPU, we deploy the benchmark on A100-40GB using the Huggingface Framework\cite{wolf-etal-2020-transformers} as the baseline GPU implementation. For the optimized GPU implementation, we accelerate the ternary linear layer with SoTA Ladder\cite{wang2024ladder} GPU kernels, which support mixed-precision operations, and optimize the attention layer with Flash Attention from XFormer\cite{xFormers2022}. 
For GLA models, we use the optimized triton kernel in FLA\cite{yang2024fla} as GPU implementations.
We measure GPU latency and on-chip power by inserting torch.cuda.synchronize with nvidia-smi API at the start and end points, and calculate the average result of repeated workloads. 
For comparison with CPU, we deploy the benchmark on PC platform with an Intel i7-12700 CPU@4.5GHz, and \texttt{\string~}30 GBps of DRAM bandwidth. We use the SoTA CPU framework, bitnet.cpp\cite{wang2025bitnet}, an enhanced version of llama.cpp\cite{GgmlorgLlamacpp2025} optimized for BitNet, to evaluate performance.

\subsection{Algorithm Performance}

\begin{table*}[!t]
\scriptsize
\begin{minipage}{0.65\linewidth}
\centering
\label{table:bitnet-result}
\caption{Perplexity and results of Sparse BitNet, BitNet and LLaMA model on the end tasks.}
\begin{tabular}{ccccccccc}
    \toprule
    Model & Size &  \textbf{PPL$\downarrow$} & \textbf{ARCc$\uparrow$} & \textbf{ARCe$\uparrow$} & \textbf{HS$\uparrow$} & \textbf{PQ$\uparrow$} & \textbf{WGe$\uparrow$} & \textbf{Avg.$\uparrow$} \\
    \midrule
    LLaMA-LLM-FP16         & \multirow{4}{*}{1.3B} & 10.82  & 27.90 & 45.16 & 47.65 & 69.91 & 53.35 &  48.79  \\
    BitNet            &                       & 11.27 & 27.65 & 45.33 & 46.86 & 68.39 & 54.06 & 48.46 \\
    BitNet+DAS        &                       & 11.32 & 27.42 & 44.16 & 46.76 & 68.39 & 54.10 & 48.15 \\
    BitNet+DAS+LPSA   &                       & 11.39 & 27.29 & 44.10 & 46.60 & 68.27 & 54.10 & 48.08 \\
    \midrule
    LLaMA-LLM-FP16             & \multirow{4}{*}{3B} & 9.61   & 29.95 & 48.11 & 55.25 & 71.76 & 57.46 & 52.51  \\
    BitNet                &                     & 9.71   & 29.27 & 49.41 & 54.42 & 70.89 & 57.54 & 52.30  \\
    BitNet+DAS            &                     & 9.97   & 28.46 & 49.50 & 54.40 & 71.16 & 56.74 & 52.05 \\
    BitNet+DAS+LPSA       &                     & 10.18  & 28.03 & 49.37 & 54.34 & 70.96 & 56.67 & 51.87  \\
    \bottomrule
    \end{tabular}
\end{minipage}%
\begin{minipage}{0.35\linewidth}
\centering
\caption{GLA Results with Various Optimizations}
\label{table:gla-result}
\begin{tabular}{cccc}

\toprule
 & & \textbf{Wiki.} & \textbf{LMB.}\\
Model & Size &  PPL$\downarrow$ & PPL$\downarrow$ \\
\midrule
GLA-FP16         & \multirow{3}{*}{340M} & 28.83 & 43.35 \\
GLA +TQ     &   & 29.10 & 43.94 \\
GLA +TQ+DAS     &  & 29.31 & 44.01 \\
\midrule
GLA-FP16       & \multirow{3}{*}{1.3B} & 17.40 &  14.51 \\
GLA+TQ     &  & 18.13 & 14.60 \\
GLA+TQ+DAS     &  & 18.30 & 15.30 \\
\bottomrule
\end{tabular}

\end{minipage}

\end{table*}

\begin{figure}[!t]
    \centering
    \includegraphics[width=8.8cm]{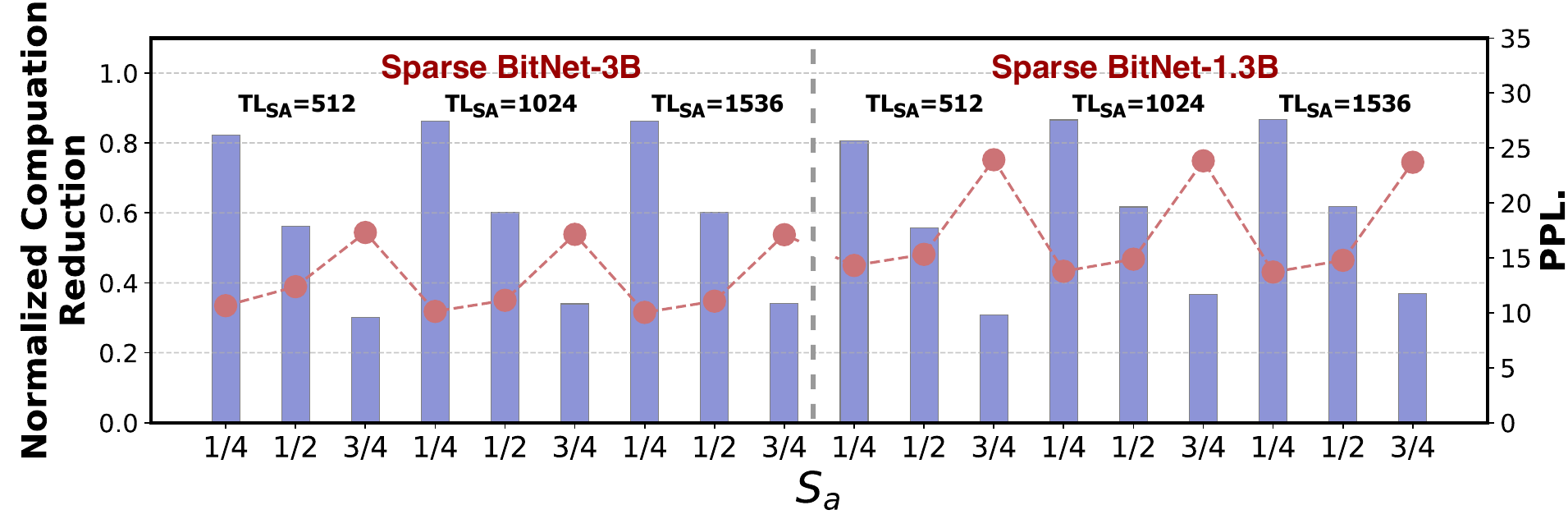}
    \vspace{-15pt}
    \caption{Sparse BitNet performance ablation over different N:M sparse ratio $S_a$ in DAS, and row-wise valid token length $TL_{SA}$ in sparse attention.}
    \vspace{-15pt}
    \label{fig:alg_ablation}
\end{figure}


\textbf{Overall  Performance:}
We evaluate the two models on seven NLP benchmarks,
including c4\cite{C4} for PLL, as well as ARC-[Challenge, Easy]\cite{ARC2019}, HellaSwag\cite{zellersHellaSwagCanMachine2019}, PIQA\cite{bisk2020piqa}, and Winogrande\cite{sakaguchi2021winogrande} for accuracy. 
As shown in Table \ref{table:bitnet-result}, compared to full-precision LLaMA with the same model size, the original dense BitNet introduces an accuracy loss of 0.33/0.2 in two models. 
Compared to dense BitNet, DAS introduces an accuracy loss of 0.31 and 0.24, but eliminates 50\% of linear layer computations.
Enabling sparse attention results in a further accuracy loss of 0.08/0.18 in 1.3B/3B models, but enable the LPSA workflow with full pipeline computation in the TENET architecture.


\textbf{Discussion on Accuracy:}\label{sec:abl}
As shown in Figure \ref{fig:alg_ablation}, we analyze the impact of the N:M sparsity ratio $S_a$ in DAS and the row-wise token length $TL_{SA}$ in sparse attention on inference perplexity (lower is better), using BitNet-3B and BitNet-1.3B on the Wikitext2\cite{wikitext} dataset. 
Notably, when $S_a$ reaches 3/4, a significant perplexity increase occurs, particularly for BitNet-1.3B model. 
In contrast, DAS with $S_a$ of 1/2 reduces approximately 40\% of the inference computation without notable performance degradation. On the other hand, increasing $TL_{SA}$ from 512 to 1536 yields marginal improvement in perplexity.
To balance hardware and algorithm efficiency, we adopt $S_a$=1/2 and $TL_{SA}$=1024 for the following hardware evaluations.
Furthermore, block-wise sparse attention mechanisms can be seamlessly integrated with LPSA workflow without incurring additional scheduling overhead.

\subsection{Hardware Performance}

\begin{figure*}[!t]
    \centering
    \includegraphics[width=18cm]{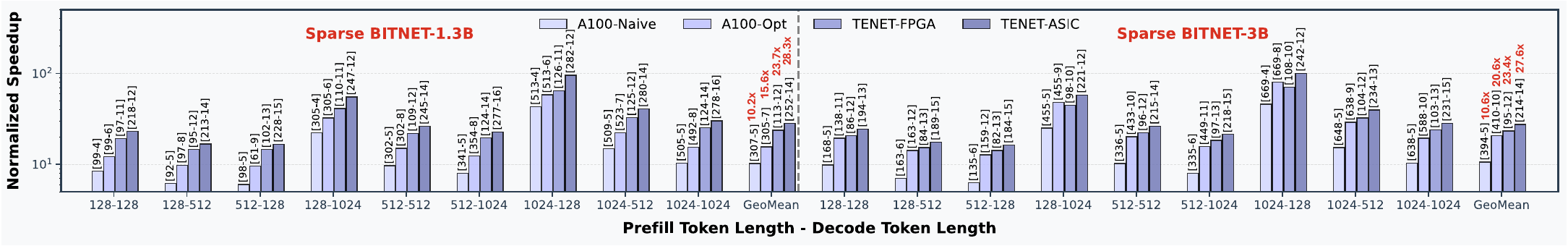}
    \vspace{-10pt}
    \caption{Speedup of TENET and A100 GPU over optimized CPU on different models and workloads (log scale); data label [a,b] refers to the speedup of the prefilling and decoding stage.}
    \label{fig:all_speedup}
    \vspace{-10pt}
\end{figure*}

\textbf{Throughput Improvement} 
Figure \ref{fig:all_speedup} shows the speedup in the prefill/decode stages of four implementations (naive and optimized A100 GPUs, TENET-FPGA, and TENET-ASIC) compared to the CPU baseline, evaluated over different workloads and two model sizes.
Despite extensive optimization, the CPU baseline remains constrained by limited compute throughput and memory bandwidth.
A100-Opt improves the attention mechanism and utilizes the A8W2 kernel, resulting in a performance improvement of 1.73x compared to A100-Naive.
TENET-FPGA achieves an end-to-end speedup of 1.51x/1.13x over A100-Opt for two model sizes, while TENET-ASIC achieves 1.8x/1.33x due to its higher compute throughput.
Although the limited on-chip computation resources result in insufficient parallelism during the prefilling stage for long $TL$ compared to the A100 GPU, TENET-FPGA and TENET-ASIC achieve average speedups of 1.45× and 1.7× over A100-Opt in the decoding stage, respectively, due to extensive optimizations in memory access.



\begin{figure*}[!t]
    \centering
    \includegraphics[width=18cm]{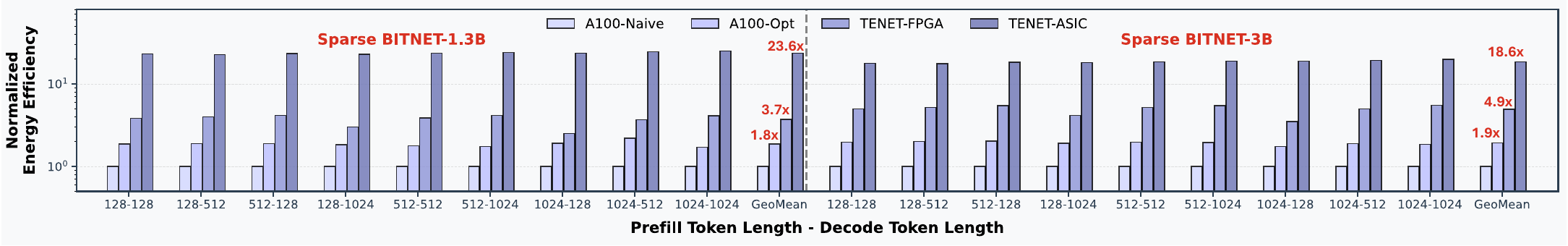}
    \vspace{-10pt}
    \caption{Energy efficiency improvement of \nickname over A100 GPU on different models and workloads (log scale). }
    \vspace{-10pt}
    \label{fig:all_energy}
\end{figure*}

\textbf{Enrergy Efficiency Improvement} Figure \ref{fig:all_energy} shows the comparison of energy efficiency between GPU and TENET. 
TENET-FPGA achieves an average energy efficiency improvement of 4.3x and 2.26x over the naive and optimized A100, respectively.
TENET-ASIC fully demonstrates the energy efficiency benefits of the TENET architecture, achieving 21.1x and 11.12x improvements over the optimized GPU through its lightweight STL Core and memory access optimizations.


\begin{figure}[!t]
    \centering
    \includegraphics[width=8.8cm]{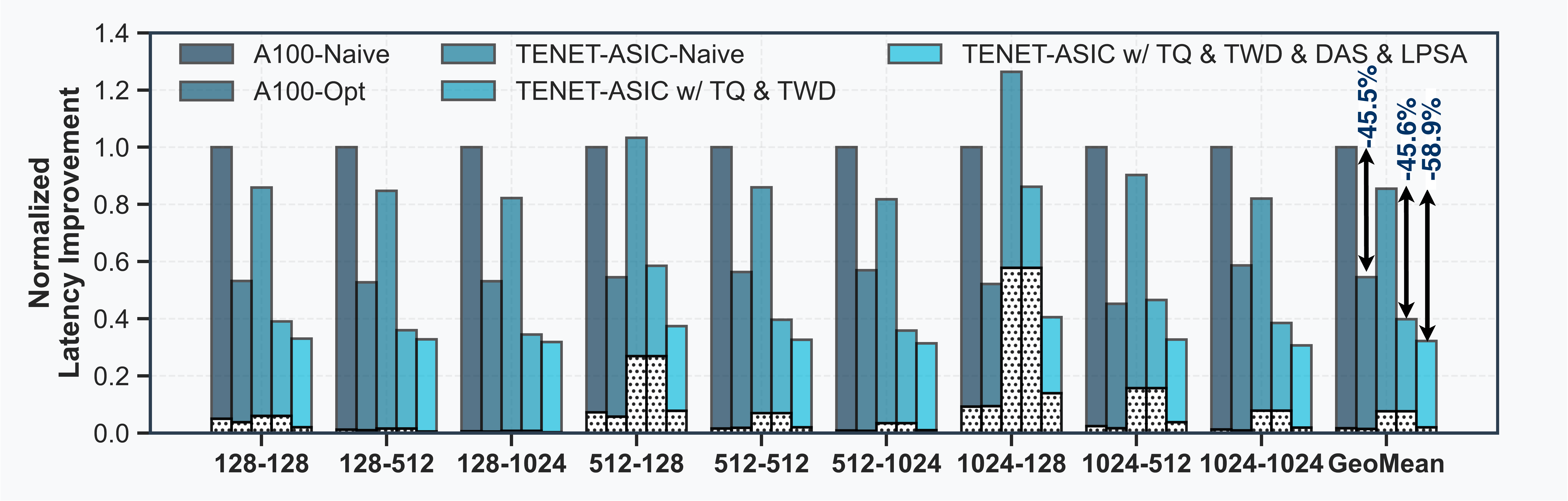}
    \vspace{-15pt}
    \caption{TENET-ASIC latency improvement over A100 GPU with different optimizations on Sparse BitNet-1.3B. The colored part corresponds to decoding and the white part corresponds to prefilling.}
    \vspace{-10pt}
    \label{fig:hardware_abl}
\end{figure}

\textbf{Performance Breakdown:} 
Figure \ref{fig:hardware_abl} shows the sparse BitNet-1.3B inference latency during the prefilling phase (colored segment) and decoding phase (white dotted segment) across A100-Naive, A100-Opt, and TENET-ASIC under different optimization levels. 
The baseline TENET-ASIC-Naive implementation assumes no optimization and stores ternary weights in INT8 format.
It achieves only 60\% of the performance of A100-Opt, due to A100's higher peak memory bandwidth for decoding and greater compute throughput for prefilling.
TWD exploits the efficiency of the ternary format and enhances compression, resulting in a 45.6\% reduction in latency.
DAS further improves on-chip throughput by reducing the hardware overhead of the HMVM core and optimizing activation memory access.
The LPSA dataflow fuses QKV projection with sparse attention at tile granularity, eliminating redundant energy-hungry DRAM traffic during prefilling, and completely hiding attention latency with the compute pipeline.
Together, DAS and LPSA bring an additional 13.3\% reduction in inference latency.
In summary, TENET-ASIC optimizes both the prefilling and decoding stages of ternary LLM inference, reducing inference latency by 67.9\% and 40.5\% compared to A100-Naive and A100-Opt, respectively.


\begin{scriptsize}
\begin{table}[t!]
  \centering
  \caption{Area and Power Breakdown of TENET-ASIC@500MHz}
  \label{table:overhead}
  \begin{tabular}{l|lcc}
    \toprule
    \textbf{Module} & \textbf{Parameter} & \textbf{Area[mm${^{2}}$]} & \textbf{ Power[mW]} \\
    \midrule
    HMVM: STL & 16 STL Cores & \multirow{2}{*}{10.24} & \multirow{2}{*}{672} \\
     Core Array & 32x64 $/$Core & & \\
    \hline 
    HMVM: HP & 4 HP Cores & \multirow{2}{*}{45.84} & \multirow{2}{*}{3315.2} \\
     Core Array & 32x64 $/$Core & & \\
    \hline \
    HMVM: & 16 Channels & \multirow{2}{*}{0.54} & \multirow{2}{*}{20.1} \\
    Others & Fan In/Out & & \\
    \hline 
    \multirow{3}{*}{Memory} & 256KB Input Buffer & \multirow{3}{*}{7.8}  & \multirow{3}{*}{401.7}  \\
    &128KB Weight Buffer & & \\
    & 1MB KV Buffer & & \\
    \hline
    SNU & 16$\times$32 Units & 11.2 & 637.9 \\
    \hline
    SFU & 16$\times$32 Units & 6.5 & 224.8 \\
    \hline
    TMI\&IO & - & 8.0 & 391.9 \\
    \hline 
    Others & - & 0.9 & 26.15 \\
    \hline
    DRAM & \multicolumn{3}{c}{HBM2, 16$\times$ HBM channels@ 2GHz} \\ 
    \hline
    \textbf{Total} & \multicolumn{3}{c}{TSMC28nm, Area=91.0mm${^{2}}$, Power=5.7W}\\ 
    \bottomrule
  \end{tabular}
\end{table}
\end{scriptsize}
\textbf{Power, Area and Energy:} 
\textbf{{\em (a) Overhead:}} Table~\ref{table:overhead} shows the area and power breakdown of TENET-ASIC accelerator.
We configure $P_L=16$, $P_H=4$ for TENET-ASIC through DSE based on algorithm analysis, performance model, and power model.
The total power consumption is 5.7W.
The STL Core accounts for 9\% of power and 13\% of area consumption, demonstrating its suitability for integration into existing hardware architectures to accelerate ternary linear operations with minimal overhead.

\begin{figure}[!t]
    \centering
    \includegraphics[width=8.8cm]{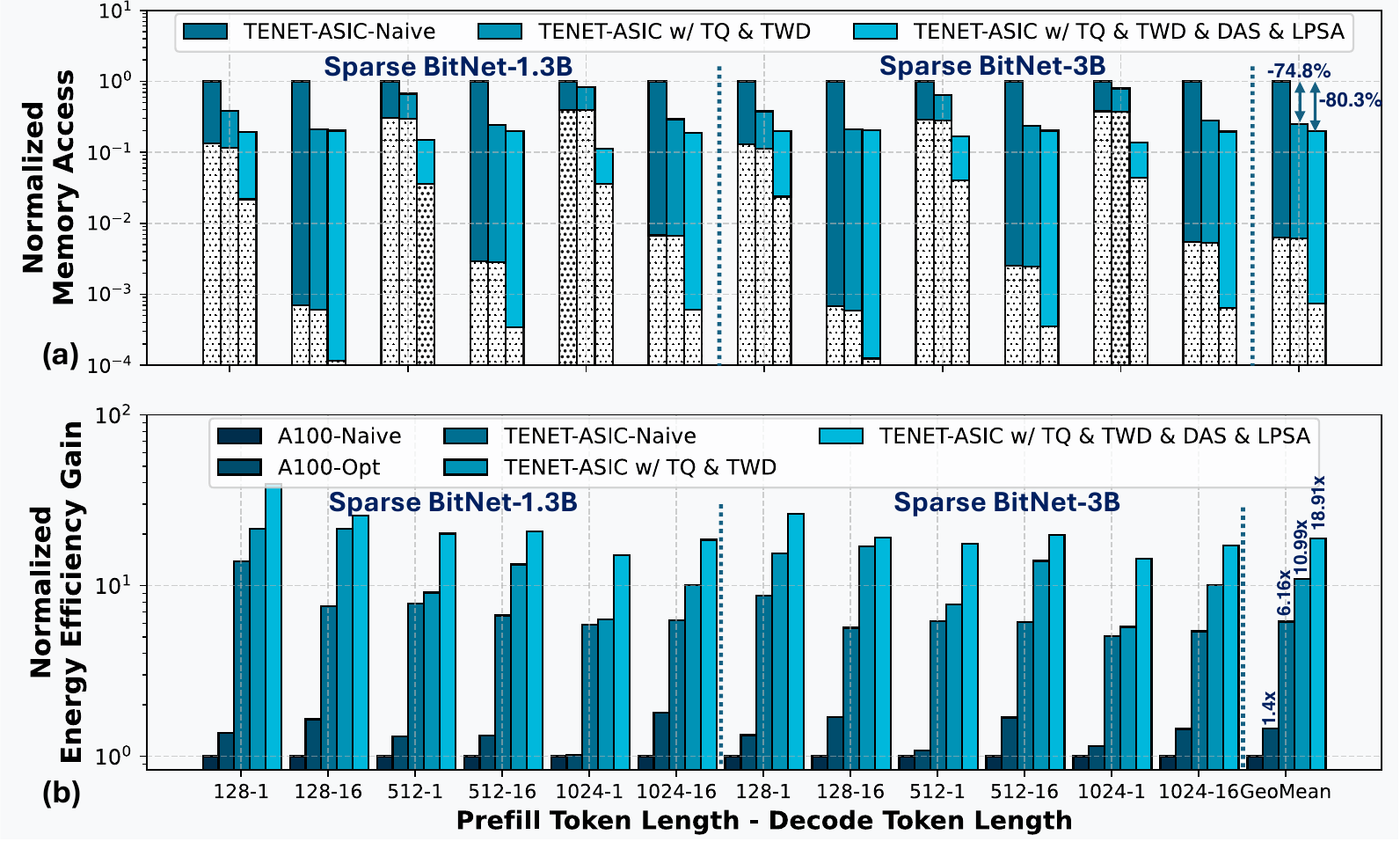}
    \vspace{-15pt}
    \caption{(a) Memory access reduction of TENET in {\em log scale}, The colored part and white part are the memory access of decoding and prefilling stage, respectively. (b) Energy efficiency of TENET-ASIC over Nvidia A100 GPU on different models and workloads}
    \vspace{-10pt}
    \label{fig:memory_ablation}
\end{figure}

\textbf{{\em (b) Memory Optimization and Energy Efficiency:}} Off-chip data access becomes a primary bottleneck in low-batch LLM inference systems. As illustrated in Figure~\ref{fig:memory_ablation}, TENET optimizes memory overhead in both decoding and prefilling stages.
During the decoding stage, the main memory access overhead comes from the model weights, TWD fully utilizes the compression opportunities of ternary weights to alleviate memory-bound issue, reducing 74.8\% memory access compared to the naive baseline on average.
During the prefilling stage, the primary source of memory access redundancy arises from frequent intermediate data transfers in the attention mechanism.
By combining DAS and LPSA, TENET effectively eliminates these redundant accesses, resulting in an 80.3\% reduction in end-to-end memory access.
As shown in Figure \ref{fig:memory_ablation}(b), TENET-ASIC with full optimizations delivers energy improvements of 18.9× and 13.5× over A100-Naive and A100-Opt, respectively.

\subsection{Linear Attention on TENET}
\begin{figure}[!t]
    \centering
    \includegraphics[width=8.8cm]{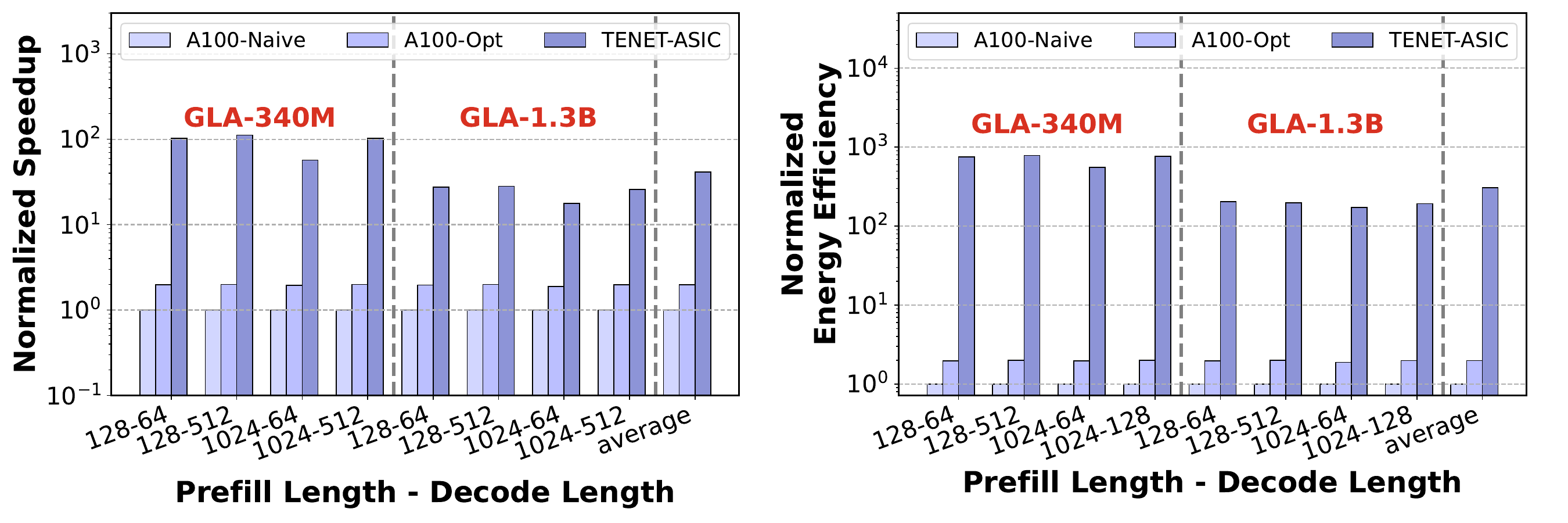}
    \vspace{-15pt}
    \caption{Speedup (left) and energy efficiency gain (right) of TENET-ASIC over A100 on GLA models}
    \label{fig:exp_gla}
    \vspace{-10pt}
\end{figure}

Table \ref{table:gla-result} shows the perplexity of GLA-340M and -1.3B on Wikitext2(Wiki.) and LAMBADA(LMB.). 
After performing ternary quantization and DAS compression on weight and activation, the average PPL of the 340M and 1.3B models increases by 0.57 and 0.85.
Figure \ref{fig:exp_gla} shows the improvements in inference speedup and energy efficiency of TENET-ASIC over A100 for the GLA model.
Despite applying FLA, where the Triton kernel fuses most operators in linear attention, A100 exhibits limited performance for low-batch GLA inference. In contrast, TENET-ASIC achieves superior hardware utilization, delivering an acceleration of 41.25x and an energy efficiency improvement of 306.6x compared to the A100.

\subsection{Comparison with SoTA accelerators}
\begin{figure}[!t]
    \centering
    \includegraphics[width=8.8cm]{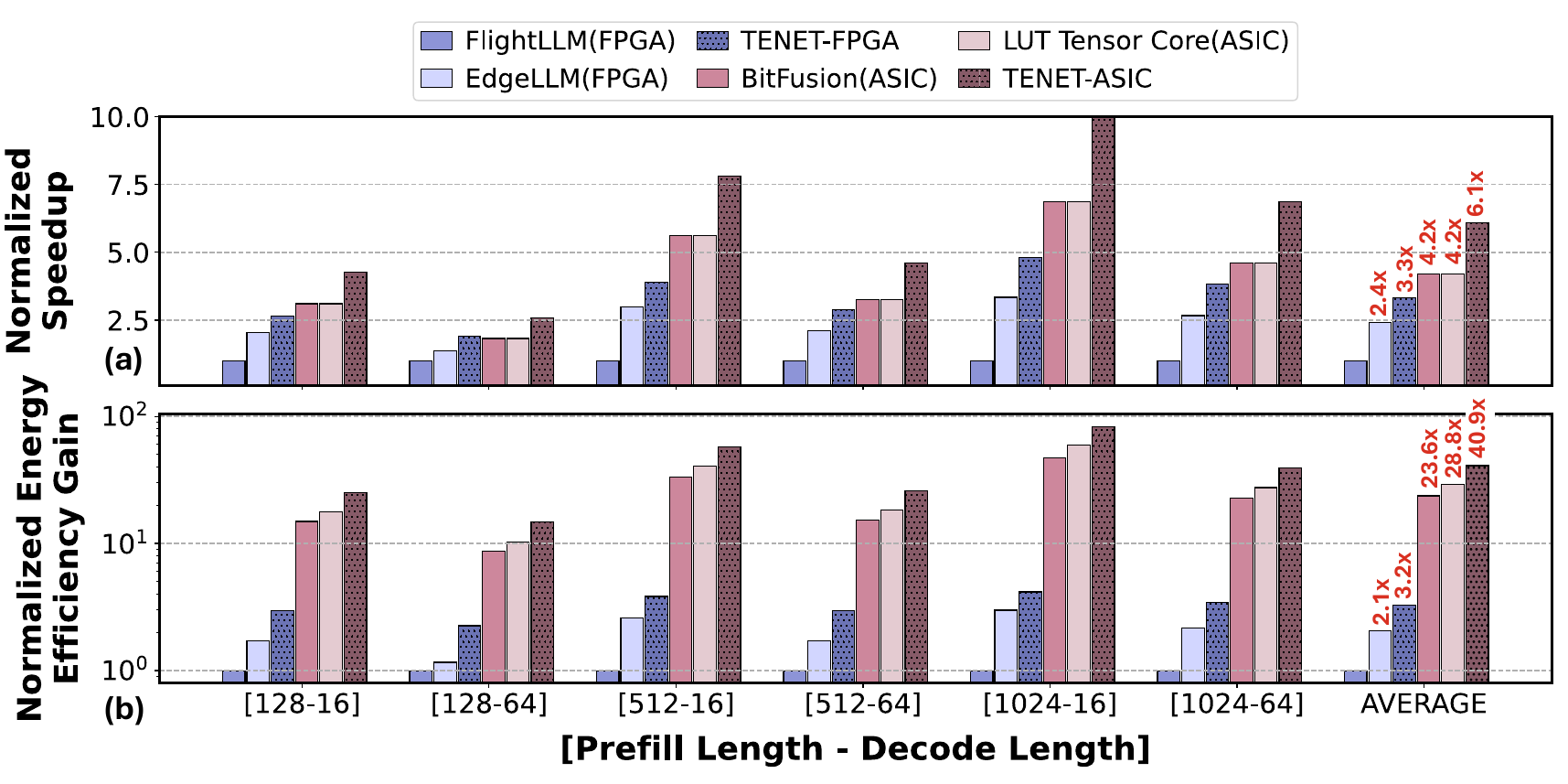}
    \vspace{-20pt}
    \caption{(a) Speedup and (b) energy efficiency gain {\em in log scale} of TENET over SoTA accelerator on FPGA and ASIC platforms with different workload on BitNet-3B.}
    \label{fig:sota_compare}
    \vspace{-10pt}
\end{figure}

Figure \ref{fig:sota_compare} shows the speedup and energy efficiency of different accelerators running BitNet-3B model. 
We build simulators based on the corresponding hardware design to evaluate their performance, achieving less than 6\% deviation using their original data.
We align the hardware parameters (clock frequency, memory bandwidth, and model precision) for fair comparison). 
To analyze the performance on the ASIC platform, we replace the HMVM engine in TENET-ASIC with BitFusion\cite{sharma2018bitfusion} Core and LUT Tensor Core\cite{moLUTTensorCore2025b} of equivalent throughput for comparison. Both alternatives support mpGEMM computations which naturally enables acceleration for BitNet.
On the FPGA platform, TENET is compared with the FPGA-based FlightLLM\cite{flightllm} and EdgeLLM\cite{edgellm}, achieving speedup of 3.3× and 1.4×, respectively. These improvements primarily stem from TENET-FPGA’s higher STL Core throughput for prefilling and its more aggressive memory compression for decoding.
On the ASIC platform, the performance advantage of TENET over BitFusion and LUT Tensor Core is mainly attributed to memory access optimizations. Through operation fusion and weight compression, TENET reduces costly off-chip memory access, achieving an average 1.49× speedup and a 1.57× energy efficiency improvement.

\section{Conclusion}

We propose \nickname, a LUT-centric architecture that unlocks the efficiency potential of ternary LLM inference on edge. 
First, we present lightweight STL core with dynamic N:M sparsity, enabling hardware-efficient ternary linear inference.
Then, we propose the TWD module to compress static ternary weights and the DPSA dataflow to reduce dynamic activation memory access. Building on top, an heterogeneous TENET accelerator is designed for end-to-end ternary LLM inference on edge. FPGA-based and ASIC-based TENET implementations deliver 4.3× and 21.1× higher energy efficiency than the A100 GPU, while reducing inference latency by 2.7×. 



\bibliographystyle{IEEEtranS}
\bibliography{refs}

\end{document}